\documentclass[letterpaper]{article}
\usepackage{graphicx}
\usepackage[utf8]{inputenc}
\usepackage[longnamesfirst]{natbib}
\usepackage{url}
\usepackage{amssymb}
\usepackage{amsmath}
\usepackage{geometry}
\graphicspath{{Plots/}}
\usepackage[section]{placeins}
\usepackage{float}
\usepackage{wrapfig}
\usepackage{appendix}
\usepackage{enumerate}
\usepackage{enumitem}
\usepackage{comment}
\usepackage{import}
\newcommand{\GG}[1]{}

\usepackage{tikz}
\usepackage{todonotes}
\usepackage{doi}
\usetikzlibrary{shapes,arrows}
\tikzstyle{block} = [rectangle, draw, 
    text width=5em, text centered, rounded corners, minimum height=4em]
\tikzset{
	>=stealth',
	point/.style={
	rectangle, rounded corners, draw=black, very thick, text width=6.5em, minimum height=2em, text centered},
	line/.style={
	->, thick, shorten <=2pt, shorten >=2pt,
	}
	}

\title{Bottoms Up: The Standard Model Effective Field Theory from a Model Perspective}
\author{Philip Bechtle\footnote{Physics Institute, University of Bonn, bechtle@physik.uni-bonn.de}, Cristin Chall\footnote{Physics Institute, University of Bonn, chall@physik.uni-bonn.de}, Martin King\footnote{Physics Institute, University of Bonn, mking@uni-bonn.de}, Michael Kr{\"a}mer\footnote{Institute for Theoretical Particle Physics and Cosmology, RWTH Aachen, mkraemer@physik.rwth-aachen.de}, Peter M{\"a}ttig\footnote{Physics Institute, University of Bonn, maettig@physik.uni-bonn.de}, \\ Michael St{\"o}ltzner\footnote{Department of Philosophy, University of South Carolina, stoeltzn@sc.edu}
}
\date{\today}

\begin{document}
\maketitle



\begin{abstract}
Experiments in particle physics have hitherto failed to produce any significant evidence for the many explicit models of physics beyond the Standard Model (BSM) that had been proposed over the past decades. 
As a result, physicists have increasingly turned to model-independent strategies as tools in searching for a wide range of possible BSM effects.
In this paper, we describe the Standard Model Effective Field Theory (SM-EFT) and analyse it in the context of the philosophical discussions about models, theories, and (bottom-up) effective field theories.
We find that while the SM-EFT is a quantum field theory, assisting experimentalists in searching for deviations from the SM, in its general form it lacks some of the characteristic features of models. 
Those features only come into play if put in by hand or prompted by empirical evidence for deviations. 
Employing different philosophical approaches to models, we argue that the case study suggests not to take a view on models that is overly permissive because it blurs the lines between the different stages of the SM-EFT research strategies and glosses over particle physicists' motivations for undertaking this bottom-up approach in the first place. 
Looking at EFTs from the perspective of modelling does not require taking a stance on some specific brand of realism or taking sides in the debate between reduction and emergence into which EFTs have recently been embedded.
\end{abstract}

\section{Introduction}

To date, the Large Hadron Collider (LHC) has produced no significant evidence in favour of the many models beyond the Standard Model (BSM) that have been previously proposed. 
The data gathered since has increasingly reduced the parameter space in which these existing models dwell, while not indicating along which lines new models should be developed.
This has led to an increase in use of approaches that do not contain any explicit assumptions about the features of BSM physics apart from its consistency with the SM and general principles of quantum field theory.
These approaches are often called `model independent' and we will describe this in more detail in Section~\ref{sec:eftintro}.
We will focus on a presently popular tool in model-independent approaches, the Standard Model effective field theory (SM-EFT), which provides a framework for systematically quantifying constraints on SM deviations.
In this paper, we examine some of the philosophical implications of this turn towards model-independence, in particular the under-emphasised practical aspects of the use of the SM-EFT at the LHC.
We want to ask: what is the SM-EFT and the bottom-up research strategy, what is model-independent about it, and how should we understand it in the context of the philosophical literature on models and effective field theories?


An EFT efficiently describes phenomena on a specific energy scale by availing itself of a separation of scales and absorbing the physics at higher energies only into the parameters of lower scale physics that are determined by experiments at those lower energies. 
The main point is that the theory can be organised in terms of different scales and that the limit of validity is already built into the theory. 
While top-down EFTs make certain assumptions about high-energy phenomena to specify or justify the low-energy theory, bottom-up EFTs are typically developed to parametrize observations or aid in searches for hitherto unknown high-energy physics.
As we shall show, EFTs share key properties with models, among them limited validity, autonomy, and practical efficiency. 
But as \citet{hartmann2001} has argued they also exhibit features that are typical for theories. 
This raises the question to what extent SM-EFT exhibits features of theories and models, and what this teaches about the relationship between models and theories in elementary particle physics.
\citet{hartmann2001} and other philosophers mostly focus on what are called \textit{top-down} EFTs, but the SM-EFT is part of a \textit{bottom-up} strategy.
Top-down EFT typically produce very specific predictions while bottom-up EFTs are rather used to obtain limits for possible physics. 
In the case of SM-EFT---or similar approaches---the bottom-up EFT is explicitly constructed as deviations from the SM with its parameters constrained by the available experimental evidence. 
We find that the answers to the questions we ask differ depending on the stage of the bottom-up strategy.

In this paper we examine the SM-EFT through the lens of the model debate, specifically through the popular models-as-mediators approach of \citet{morganmorrison}, which was generalised by \citet{mccoymassimi}, as well as the semantic approach of \citet{hartmann1999} and the artefactual account of \citet{knuuttila2011}. 
In Section~\ref{sec:data}, we provide empirical data indicating the increasing popularity EFT approaches in particle physics. 
In Section~\ref{sec:eftintro}, we introduce three kinds of approaches presently applied by particle physicists in the search for new physics: full BSM models, simplified models (of BSM physics)\footnote{We will follow the terminology common among particle physicists and refer to ‘simplified models’ as a particular kind of simplified BSM model, rather than as some model that has been simplified. The precise notion of a simplified model will become clearer in Section~\ref{sec:SimpMod}}, and SM-EFT. 
Section~\ref{sec:models} provides the relevant background from the contemporary philosophical debates about models and effective field theories. 
Section~\ref{sec:classification} presents the model-independent strategy for finding new physics using the bottom-up approach of the SM-EFT, which we subdivide into three stages.
We show this strategy in action with a case study of b-physics in Section~\ref{sec:BphysicsConcepts}. 
Finally, Section~\ref{sec:analysis} provides a philosophical analysis of the SM-EFT strategy.  
In particular, we distil four lessons that concern: i) the relation that SM-EFT bears to models and theories in semantic and practice-based approaches; ii) the role of representation in models of new physics; and iii) the limits for something be considered a model in physics; iv) how a model-focused look at the presently popular practice of SM-EFT can bring a new perspective into recent philosophical debates about EFT that have taken it simply as a well-formulated theory on a particular scale that allows one to discuss traditional issues, such as realism vs. instrumentalism and reduction vs. emergence, in the light of the separation of scales that we find in fundamental physics. It also gives due space to the fact that bottom-up EFTs are part of model-independent experimental search strategies for BSM physics.

\section{Diagnosing a Trend} \label{sec:data}
Before the LHC became operational in 2010, there were strong expectations for new physics and accordingly there existed a whole landscape of BSM models. 
However, none of the searches for new physics turned up anything significant enough to indicate BSM physics.
The Higgs boson was discovered, but all studies indicate that it has only SM properties \citep[see][for a detailed discussion of the discovery]{challetal1}.
Analyses are increasingly constraining the existing BSM models, and designing experimental searches based on model-specific predictions seems less and less fruitful. 
Physicists are thus increasingly turning to alternative methods for searching for new physics, they are using simplified models and so-called model-independent approaches, among them EFTs.
This trend away from model-guided searches and towards EFTs can already be documented with simple bibliographic means.
As an example, Figure~1 shows the results of a keyword search that compares some of the most popular BSM models in the Higgs sector with model-independent approach of the Higgs EFT and precision measurements.  
The papers plotted employ supersymmetric Higgs models, composite Higgs models, extended Higgs sector models that are non-supersymmetric, and the Higgs EFT approach. 

The search was conducted on the popular physics preprint archive, ArXiv.org.
Specifically, it was conducted on the phenomenology archive HEP-PH, the most relevant to models in high-energy physics.
We have shown in previous publications that such keyword searches provide a good first look at the dynamics of particle physics models and that the corresponding trends can be confirmed by expert interviews.\footnote{For a full description of the methods used, see \citep{challetal1}.}
The main reason such figures provide a reasonable first look is that the mentioned preprint archives are the main scientific venue for particle physicists.
Additionally, ArXiv.org features an automatic keyword tagging system.
This system provides us with a simple method of gauging some of the trends in particle physics by seeing how the numbers of papers with different keywords change over time.\footnote{The keywords used were the following. For EFT: find k "Higgs" and (k "effective field theory" or k "precision measurement"). For SUSY: find (k "supersymmetry"or k "minimal supersymmetric standard model" or k "MSSM") and k "Higgs". For Composite Higgs: find (k "technicolor" or k “Higgs particle: composite" or k "Higgs particle: Goldstone particle" or k "pNGB" or k "top: condensation" or k "little Higgs model" or k "dynamical symmetry breaking") and k "Higgs" and not (k "supersymmetry" or k "minimal supersymmetric standard model" or k "effective field theory"). For Non-SUSY extended Higgs: find (k "Higgs particle: doublet: 2" or k "2HDM" or k "Higgs particle: triplet" or k "Higgs particle: doublet: 3" or k "Higgs particle: charged particle") and not (k "supersymmetry*" or k "minimal supersymmetric standard model" or k "effective field theory").}
The keyword searches were conducted in such a way as to minimize overlap between the terms by specifying, for instance, that the composite Higgs papers include neither supersymmetric models nor an EFT approach. 
\begin{figure}
\includegraphics[width=0.9\textwidth]{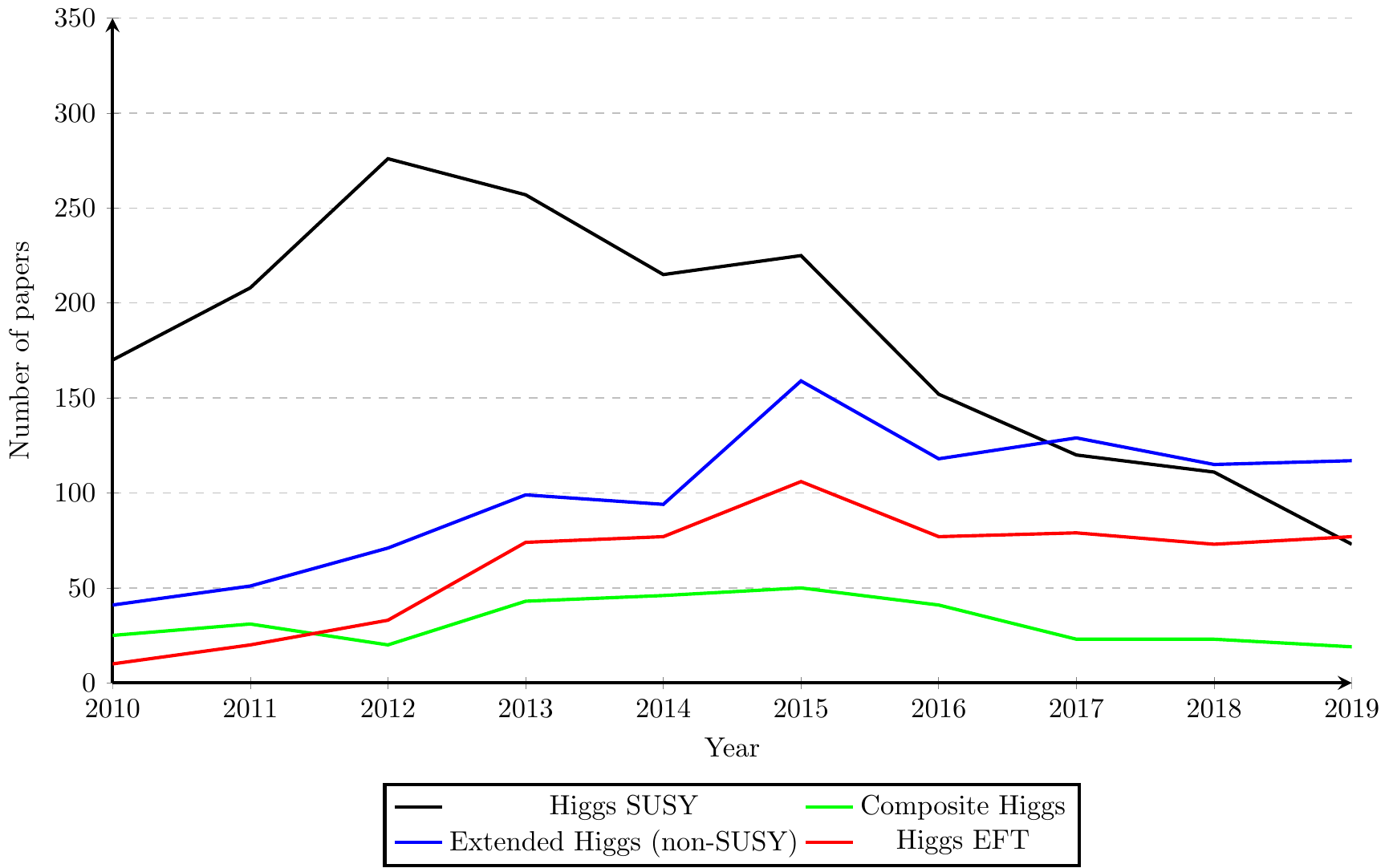}
\caption{The number of papers per year for the most popular BSM models in the Higgs sector and the Higgs EFT.} 
\label{fig:lineplot}
\end{figure}

\begin{figure}
\includegraphics[width=0.9\textwidth]{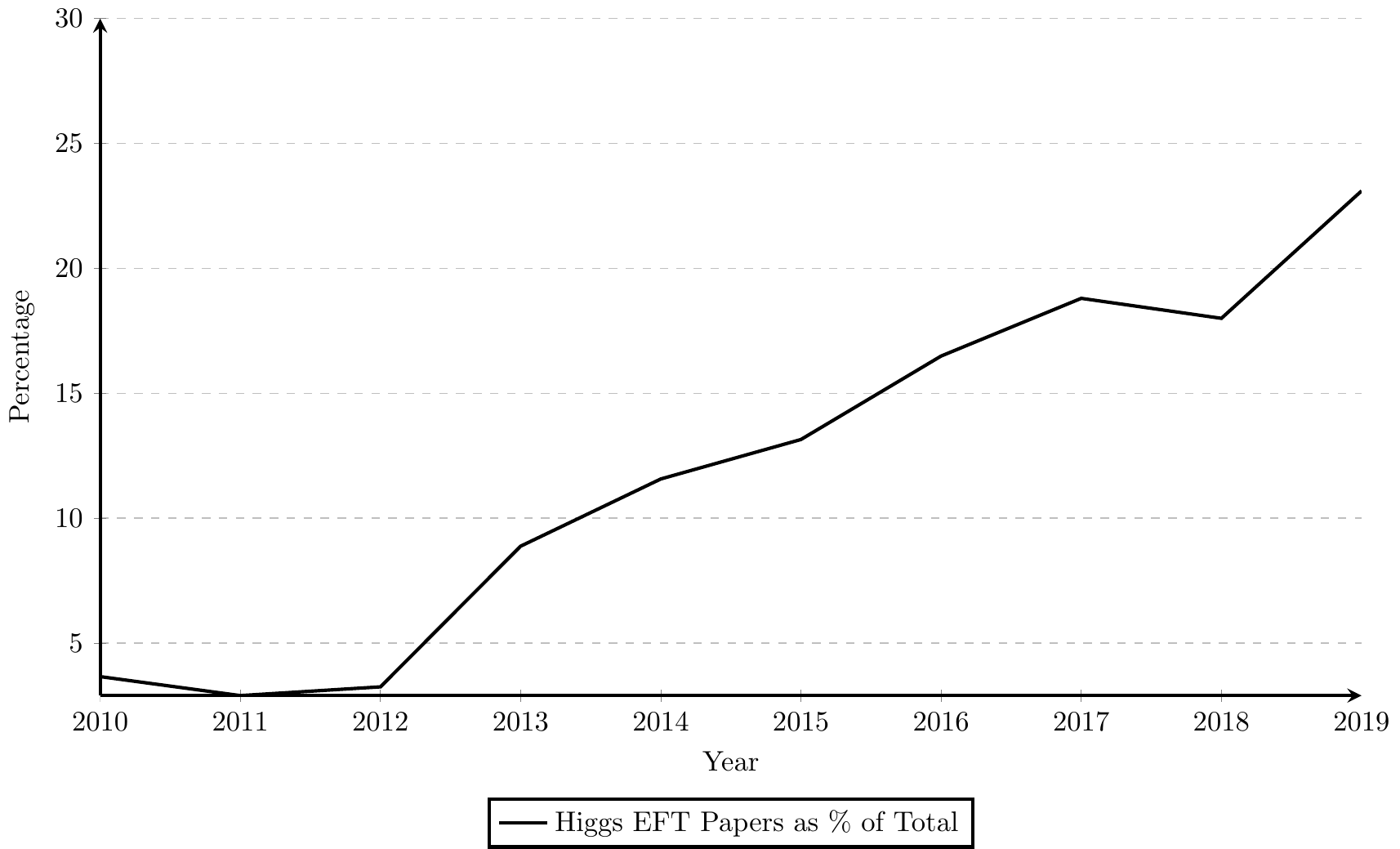}
\caption{The total number of papers from the most popular models of BSM in the Higgs sector (sum from Fig.~\ref{fig:lineplot}) against the number of papers on Higgs EFT as a percentage of that total.}
\label{fig:percent}
\end{figure}

One can see in Fig.~\ref{fig:lineplot} a flurry of model building up to and past the Higgs discovery, lasting until 2015. 
Since then, there has been an overall decrease in the number of papers on Higgs models, in particular a strong decrease in supersymmetric models. 
The number of non-supersymmetrically-extended Higgs models is growing, but more slowly than the number of papers of supersymmetric models is declining, hence the decreasing total. 
It should be noted that this is not one particular model, but captures many different minimal extensions, such as non-SUSY 2-Higgs doublet models, and includes those that merely search for charged Higgses.

What is of most interest to us in this paper is that the number of papers on the Higgs EFT approach generally climbs over the past decade. 
As a result, while the total number of papers is in a downtrend, there is a strongly increasing percentage of these papers that employs the Higgs EFT approach. 
The increase in the percentage of Higgs EFT papers relative to the total plotted in Fig.~\ref{fig:percent} is quite dramatic.
In 2010, only 3.6 percent of papers in the Higgs sector used EFTs, but by 2019 that has risen to more than 23 percent.\footnote{The percentage is calculated against the total number of papers as the sum of the main relevant BSM models included in Fig.~\ref{fig:lineplot}.}
The number of papers on models that introduce new fields and new symmetries declines as the number grows for those that take a model-independent approach with no such commitments. 
Of course, these plots only cover the Higgs sector and the Higgs sector is only a part of the research of high energy physics, but this is indicative of the trend in the field at large. 
We do not claim that this data is exhaustive, but take this data as a jumping-off point that shows that there is a trend that is worth the attention of philosophers.
We also take this to indicate that there is a real distinction in physics practice between the model-based and model-independent approaches.  
Part of our aim in the paper is to characterise this distinction and investigate how one should understand this trend in the context of the contemporary debates on the nature of models, theories, and EFTs.

\section{Effective Field Theories - Simplified Models - Concrete Models}
\label{sec:eftintro}

The Standard Model (SM) is the best currently available description of
fundamental interactions at the microscopic scale.\footnote{~Among the seminal papers are \citep{Weinberg:1967tq,Glashow:1961tr,Salammodel,Fritzsch:1973pi, Higgs:1964ia,Higgs:1964pj,Englert:1964et,Guralnik:1964eu,Hooft:1972fi}.}
It has been tested at experiments up to TeV energies, where the tests at the highest energies mainly come from the Large Hadron Collider (LHC). 
Within expected statistical fluctuations, the ensemble of these tests
supports the predictions of the SM to high precision.
Because the SM has a variety of internal and external deficiencies, it is expected that the SM is connected to a yet-unknown, more fundamental theory with new degrees of freedom at an energy $\Lambda$, somewhere between the TeV-scale tested at the LHC and the Planck scale, where gravity has to be accounted for. 
If $\Lambda$ is within the energy reach of the LHC, the BSM physics
will manifest itself in the form of new resonances and new or modified
interactions. If $\Lambda$ is above, but not too far above the TeV
scale, effects of BSM physics could still manifest themselves through
new effective interactions between SM fields. In the absence of any
clear deviations of experimental data from the SM predictions,
particle physics has developed several strategies to theoretically
approach BSM physics that will be discussed in this section:
expanding the SM into a SM-EFT in a bottom-up approach; a top-down
approach based on concrete BSM models; and a mixture of top-down and
bottom-up elements realized in simplified models.

Given the outstanding success of the SM, these approaches embed the SM, so that the Lagrangian 
for new physics can be schematically written as 
\begin{equation}
{\cal L}_{\rm new} = {\cal L}_{\rm SM} + {\cal L}_{\rm BSM}.
\label{eq:BSM}
\end{equation}
The different approaches to BSM physics differ in how ${\cal L}_{\rm BSM}$ is 
conceptualized.\footnote{Terms that include both SM and BSM fields are included in ${\cal L}_{\rm BSM}$.}
We will briefly summarize them in the following, with special emphasis on the SM-EFT.

\subsection{BSM models}

During the past decades, BSM searches at the LHC and elsewhere
were mostly guided by models that postulate specific new concepts to solve open issues of the SM, e.g.\
new symmetries, new interactions, or new spatial dimensions at some energy scale $\Lambda $. 
These concepts are reflected in ${\cal {L}}_{\rm BSM}$ 
by at least explicitly adding new (non-SM) fields
and their interactions.
No assumption is made on their UV completeness.
Prominent examples of such models are supersymmetry~\citep{Wess:1974tw}, a new
symmetry between fermions and bosons\footnote{Throughout the paper the term `supersymmetry'
denotes a field theory with the SM supplemented by the additional supersymmetry.}, technicolor
models~\citep{Weinberg:1975gm,Susskind:1978ms}  which assume a new strong interaction, and
the Randall Sundrum model~\citep{Randall:1999ee} with an extra spatial dimension.  
These new concepts might be realized in a fundamental theory at (much) higher energies than the SM. However, in a top-down approach
observable deviations from the SM expectations at LHC energies, like new particles or 
modifications of cross sections, can be deduced in these models. 

Consider, for example, the case of supersymmetry (SUSY).
Supersymmetry is the largest possible extension of the Poincar\'{e}
group. This concept
leads to the idea of a symmetry between fermions and
bosons. All the SM particles have supersymmetric partners with spins
differing by $1/2$. 
TeV-scale SUSY models have typically about 100 new parameters and could address many outstanding issues of the SM by providing gauge coupling unification, describing the nature of dark matter, and solving the hierarchy problem.
However, as of yet, no signal of SUSY has been found for any of the available models.

\subsection{Simplified Models}
\label{sec:SimpMod}

Due to the complexity of full BSM models, systematic simplifications have been developed in recent years that facilitate
searches for BSM physics. 
The main idea is not to use a whole,
concrete BSM model, but focus on one or a few of its new
features~\citep[see, e.g.,][]{alwall2009,alves2011}. To achieve this
formally, one may consider scenarios where almost all the new particles of
a BSM model have masses larger than the energies probed at the current
collider. Only a few selected new particles with TeV-scale masses are
then treated as relevant for experimental and phenomenological considerations.

A prominent example is the decay of a supersymmetric top quark
$\tilde{t}$. In a complete SUSY framework, it can decay into a large
number of final states. Such a decay can be described by several BSM models.
In a simplified model, one restricts, for example, the
decays to $\tilde{t}\rightarrow t\chi^0$ (see e.g.~\cite{Aaboud:2017ayj}), while all other particles
are assumed to be too heavy to be produced in the decay.  
Here $\chi^0$ is the lightest neutral supersymmetric particle and considered a candidate for 
dark matter since it interacts at most very weakly with normal matter.  
Such a
simplification has repercussions on the interpretation of a
measurement.  Restricting SUSY to just one (or two) particles takes
these out of the context of a specific model and just provides an
experimental signature without any relation to other processes in the
model. Here, the simplified model is motivated in a top-down way from
the supersymmetric model. While such a signature may be interpreted as a signal for
$\tilde{t}$, it can equally well come from other objects in other BSM
models with an equivalent decay, e.g. scalar third generation
leptoquarks that decay into $t+\bar{\nu}_t$~\citep{Aad:2015caa} or a generic Dark Matter particle with affinity to the
top quark. In this spirit, simplified models can also be constructed
in a bottom-up way: by building final states comprised of
combinations of new particles with different quantum numbers,
independent of any theory predicting these particles and their interactions.

\subsection{SM-EFT}

Effective field theories (EFTs) are of central importance in particle
physics \citep[see, e.g.,][]{Georgi:1994qn}. 
In this subsection, we will give an overview of their use and development in particle physics and introduce a recent prominent case in the search for deviations from the SM.
\footnote{While the notion of EFT is well-established in contemporary particle physics, some authors, among them \citet{wells2012} and \citet{Rivat2020-RIVPFO} consider them a continuation of previous research strategies, ranging from effective extensions of classical mechanics to the S-matrix theory and other black-box models in particle physics. Such a broader historical analysis is beyond the scope of the present paper. The Fermi theory discussed in this section fulfils all the characteristics of present EFTs, including the separation of scales, and is often cited as an archetypical EFT by particle physicists.}

\subsubsection{A historical blue-print: weak interactions}
\label{ssec:Fermi}

The Fermi theory of weak interactions~\citep{Fermi1934}, which aimed to describe the nuclear $\beta $-decay
as a pointlike interaction of four fermions (a proton, neutron, electron and a neutrino),
may serve as successful application of  a bottom-up approach to new physics using EFT procedures,
although, when Fermi developed his formalism in 1933, the concept of a
Quantum Field Theory and hence the concept of an effective QFT was not
yet established. Therefore for the sake of this discussion it is useful to translate Fermi's Ansatz
into the language of a modern QFT.
Here, we also replace neutrons and protons by quarks and include the 
helicity structure of the interaction. Then, the Fermi-interaction can
be written as
\begin{equation}\label{eq:Fermi}
{\cal L}_{\rm Fermi} = -\frac{G_F}{\sqrt2}\bar\nu\gamma^\mu(1-\gamma^5)e\,\bar q\gamma_\mu(1-\gamma^5)q'\,.
\end{equation}
The quark, electron and neutrino field operators are denoted by $q,\ q',\ e$ and $\nu$, respectively, and $\gamma^\mu, \gamma^5$ are the Dirac gamma matrices. 
The strength of the interaction is set by the Fermi constant $G_F = 1.15 \times 10^{-6}$~GeV$^{-2}$. 
In the decades following Fermi's approach to $\beta $-decay, the formalism  was successfully applied to the decay of the muon, and  to transitions between quarks. 
Fermi theory allowed one to include parity-violating couplings of weak interactions after its discovery in nuclear transitions, to understand special parameters to describe transitions between hadrons (CKM matrix), and serves to measure the mass of $\nu_e$. 
Thus, beyond describing just one process, the Fermi theory was a way to describe all weak interaction processes at low energy.
As a consequence, the formalism of Fermi became the standard representation of low energy weak interactions. 
The Fermi theory has all features of an EFT. 
It allows one to efficiently compute the production rates of weak interactions and explicitly specifies the scale at which the theory breaks down.
As the Fermi coupling $G_F$ is dimensionful, production rates
calculated within the Fermi theory will violate the probabilistic
interpretation of quantum theory (`unitarity violation') at
some energy $\Lambda \gg E_{\mathrm{\beta-decay}}$. For the Fermi
theory, this unitarity violation occurs at a centre-of-mass energy of
$E_{\rm unit.\, viol.}\approx \sqrt{1/G_F} \approx
300\,\mathrm{GeV}\gg E_{\mathrm{\beta-decay}}$. Thus, at energies near
$E_{\rm unit.\, viol.}$ new phenomena will occur and the Fermi theory
has to be replaced by a more fundamental theory. 

\begin{figure}[htbp]
\begin{center}
\includegraphics[width=0.4\textwidth]{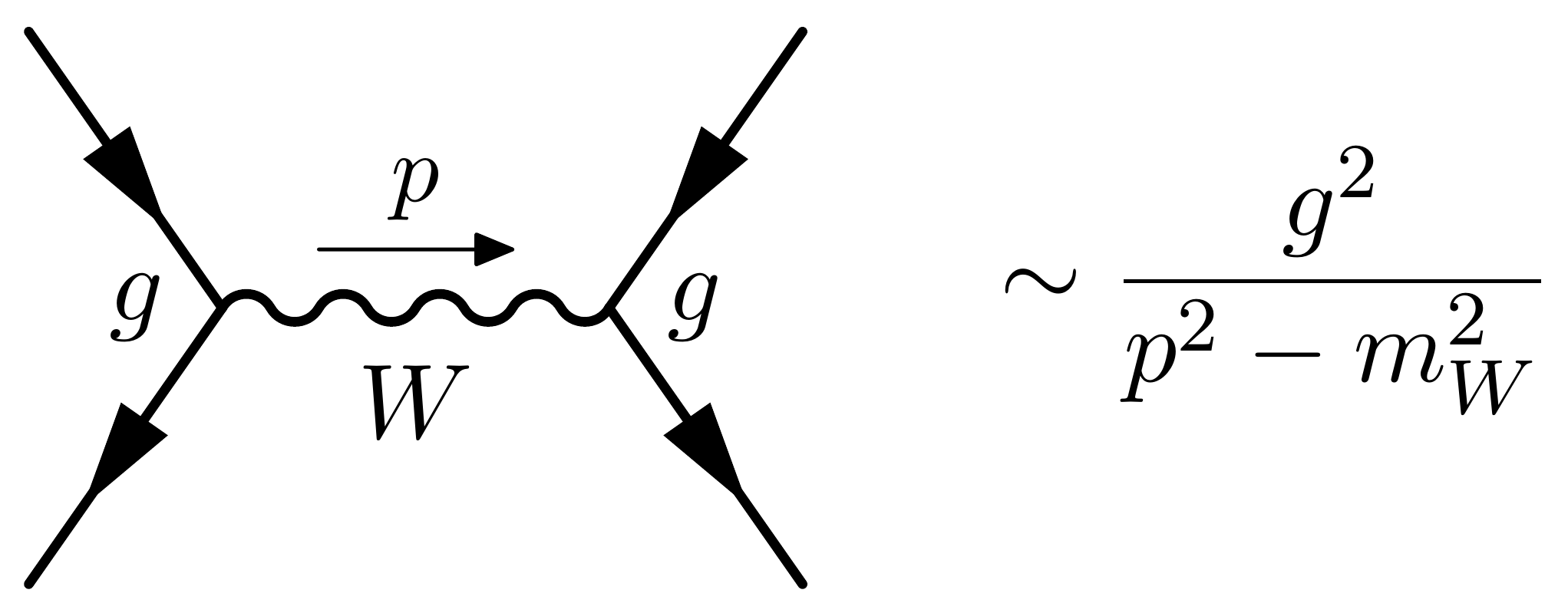}\qquad\qquad
\includegraphics[width=0.4\textwidth]{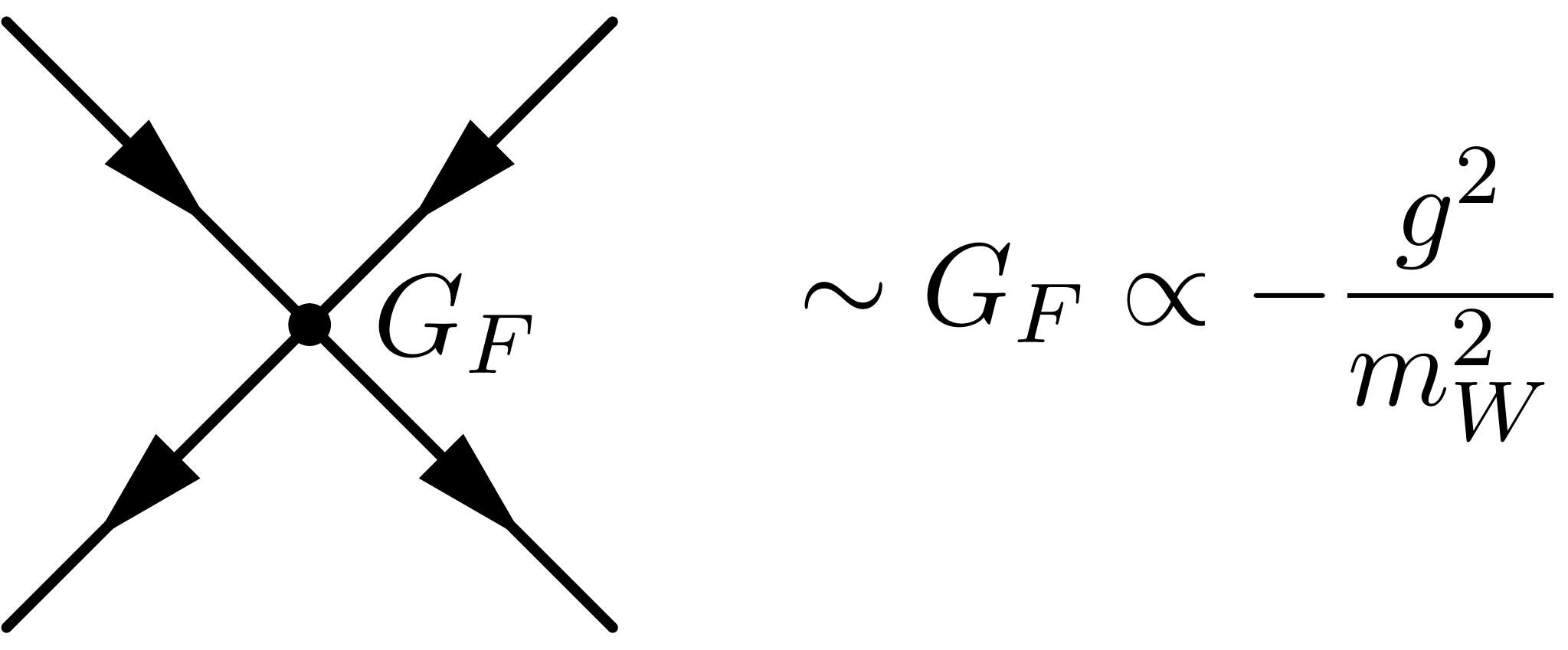}
\caption{Weak interactions as described in the Standard Model (left) and in Fermi's four-fermion theory (right).}
\label{fig:weak}
\end{center}
\end{figure}

These phenomena are the existence of the $W$ and $Z$ bosons and the
electroweak interaction in the SM.
The Fermi theory is a low-energy limit of the electroweak theory of the SM, where interactions among 
fermions are mediated by the exchange of a virtual $W$ boson
with mass $m_W$ of about 80 GeV, and a 
coupling strength $g$ (see Fig.~\ref{fig:weak}, left). 
By including a $W$ propagator, the production rate is damped and unitarity violation is avoided.
For momenta small compared to the mass of the $W$-boson, $p^2\ll m_W^2$, the $W$-propagator can be expanded in powers of $p^2/m_W^2$:
\begin{equation}
\frac{g^2}{p^2 - m_W^2} = -\frac{g^2}{m_W^2}\left(1 + \frac{p^2}{m_W^2}+{\cal O}\left(\frac{p^4}{m_W^4}\right)\right) \approx -\frac{g^2}{m_W^2} \equiv -\frac{G_F}{\sqrt2}\,.
\label{eq:wprop}
\end{equation}

Thus, the SM becomes equivalent to a pointlike effective four-fermion interaction with a strength set by the Fermi constant $G_F$ (Fig.~\ref{fig:weak}, right). 
This is an example of a \textit{top-down approach}, where the EFTs can be derived from a more fundamental theory formulated at higher energy scales.
From a modern perspective---i.e., assuming the availability of the formalism of QFT but not the SM---the Fermi theory could be regarded as an
early example for a \textit{bottom-up} EFT. 
On the other hand, the SM could not be unambiguously inferred from the Fermi theory, and neither
could the energy scale at which the EFT description breaks down, since $G_F$ depends 
on both the new scale and the unknown coupling $g$.

\subsubsection{The Standard Model Effective Field Theory} \label{sec:smeftphysics}

In analogy to the Fermi theory, new effective interactions could be added
to the SM to describe a yet unknown theory of physics beyond the SM at
an energy scale $\Lambda$. In case of the Fermi theory, the new scale
was set by the $W$ boson mass, $\Lambda = m_W \approx 80$\,GeV. Since the structure and the
scale of physics beyond the
SM is not known, physicists adopt a
general bottom-up EFT to systematically evaluate potential BSM
effects at an energy scale $E_{\rm LHC}$  relevant for LHC physics. In the case of the
so-called 
Standard Model Effective Field Theory (SM-EFT)\footnote{See \cite{Buchmuller:1985jz,Grzadkowski:2010es} and the recent review
\cite{Brivio:2017vri}}, one assumes that the
new physics respects the SM gauge symmetries and that the scales are
well separated, $\Lambda\gg E_{\rm LHC}$. Note, however, that the
scale $E_{\rm LHC}$ depends on the physics process under
consideration. 

The SM-EFT describes a different phenomenology than the simplified
models or specific BSM models described above, where it is assumed that the masses of the new
particles and the experimental energy scale $E_{\rm LHC}$ are of the same
order of magnitude. In particular, this means that the SM-EFT does not
explicitly include new particles as dynamical degrees of freedom, but
assumes that any new particles would be too heavy to be directly produced in LHC experiments. 

The SM-EFT combines the SM fields to form all gauge invariant operators
${\cal O}_i^{(D)}$ with mass dimension $D$ and expands the SM
Lagrangian to
\begin{equation}
{\cal L}_{\rm SM-EFT} = {\cal L}_{\rm SM} + \sum_i c_i^{(6)} {\cal O}_i^{(6)} + \sum_j c_j^{(8)} {\cal O}_j^{(8)} + \ldots
\label{eq:smeft}
\end{equation}
The corresponding strength of the interaction is denoted by
$c_i^{(D)}$, where $c_i^{(D)} \ = \ (g_i^D/\Lambda^{D-4})$.\footnote{
  As the Lagrangian should have mass dimension four, such that the
  action $S = \int d^4x {\cal L}$ has mass dimension zero (in natural
  units $\hbar =c =1$), the coefficients $c_i^{(D)}$ must have mass
  dimension $4-D$.}  In the lowest relevant dimension at the LHC,\footnote{Note that 
dimension 5 operators include interactions of neutrinos that, given the experimental constraints
on the $c_i$, make them unobservable at the LHC.
The same is true for dimension 6 operators violating baryon or lepton number conservation. 
These are not included in Eq.~\ref{eq:smeft}.}  D = 6, the
SM-EFT contains 2499 independent operators ${\cal O}_i^{(6)}$ as well
as 2499 $c_i^{(6)}$. It is important to note that there is a freedom
of choice of which set of independent operators 
is chosen. Each independent set is called a \emph{basis}. Depending on
the choice of basis, the SM-EFT predicts different sets of 
experimental observables to be theoretically connected through the same
operators~\citep[see, e.g.,][]{Falkowski:2015wza}. Consequences of
this freedom will be further discussed in
Section~\ref{sec:classification}. 

It should be noted
that the SM-EFT expansion, eq.~(\ref{eq:smeft}), is not an expansion in $\Lambda$
itself, but in $c_i=g_i/\Lambda$. 
In consequence, experimental constraints on deviations from the SM can
only be provided in terms of $g_i/\Lambda$.
As already discussed in the framework of the Fermi theory, one
does not immediately know the (bounds on the) scale of new physics, 
for which one would have to know $g_i$.
However, there are general constraints on the theoretical applicability of SM-EFT.
Since new resonances lead to significant contributions that cannot be captured by
the SM-EFT, the SM-EFT fails to describe physics distributions close to such a new resonance.
While in principle this makes the range of validity of the SM-EFT model dependent, 
it is generally assumed that $g \sim {\cal{O}}(1)$, such that the effects of the new 
resonance(s) with mass ${\cal{O}}(\Lambda)$ become visible already at
scales much smaller than $\Lambda$; thus one requires the SM-EFT to be valid at 
$\Lambda\gg E_{\rm LHC}$. 
For our further discussion we will assume such a strong separation of scales.
The intricacies of the range of validity of the SM-EFT are discussed,
for example, in~\cite{Contino:2016jqw}.

As mentioned above, this formalism is equivalent to what has been discussed for the Fermi theory. 
In fact some of the operators in Eq.~\ref{eq:smeft} are formally identical to the one of Eq.~\ref{eq:Fermi}. 
As in the Fermi theory the operators in the SM-EFT violate
unitarity, albeit at a scale of ${\cal O}(\gg \mathrm{TeV})$ (depending on the
parameter choices) instead of $\approx
300$\,GeV for the Fermi theory. Thus again, the SM-EFT is only an approximation and
involves two distinct mass scales, the one of the SM and a much larger scale of BSM physics.

\subsection{A Brief on the Physicists' Attitude}
\label{sec:physrepres}

The physicists' turn towards using the general SM-EFT formalism is borne out of the conceived dead end of expressing ideas for new physics in terms of concrete models.
This attitude can be characterised as such: 
``In absence of a direct observation of new states [i.e.{} particles], our ignorance of the EWSB sector can be parametrized 
in terms of an effective Lagrangian''
\citep[p.~1]{Contino:2013kra}.
It is beyond this paper to discuss in depth the epistemic and pragmatic goals of physicists leading to such a characterisation.
Suffice it to note, especially in view of the following discussions, that the bottom-up SM-EFT approach is considered as being ignorant with respect to the nature of a `new state'.

What we consider important here is that a new state would have
definite properties to be scrutinized experimentally and theoretically, which is not the case for the 
general SM-EFT approach.
If an experiment would find a new state, it would have definite properties that have to be represented in theory, e.g.\ by a field $\chi $.
Vice versa, if theory would propose a new state, its properties would largely be fixed and
would provide a clear target for experiments to search for.
In this case the field $\chi $ would be a representation of a new, but
not yet observed, particle.
In contrast, as will be detailed in Sect.~\ref{sec:classification}, the EFT approach
does not start from definite properties and thus does not provide
a clear target for experimentation.  
In contrast to SM-EFT, concrete BSM models and simplified models
do assume fields or vertices representing entities and processes, and thus offer a well defined search strategy. 
It is such BSM theories that physicists ultimately strive for (also for other reasons).
This does not render EFTs useless, but 
the ``main motivation to use this framework is that the constraints on the EFT parameters can be
later re-interpreted as constraints on masses and couplings of new particles in many BSM theories,'' \citep[p.~305]{deFlorian:2016spz}.
Thus, while using EFT as a tool to parametrise constraints due to the (non)-observation of deviations
from the SM, most physicists want to reach beyond and arrive at a consistent and concrete BSM model.
The concrete BSM models and the simplified models provide targets to search for. 
If no BSM target is defined by a model, physicists characterise their search for BSM models as `model independent'. 
As will be discussed in Section~\ref{sec:classification}, the SM-EFT is considered a tool to support model-independent searches.

\section{EFTs: Between Models and Theories} 		
\label{sec:models} 

Philosophical approaches to scientific models have traditionally been connected to scientific theories. 
As a consequence, philosophers' understanding of models was often influenced by their conception of theory.
Early syntactic accounts, most prominently by logical empiricists, treated theories as axiomatic formal systems \citep[such as][]{carnap59,hempel66}
within which particular models can be articulated. 
An axiom could have only one intended model (up to isomorphism) or plenty. 
This view was criticised in part for not being an appropriate way to cast actual scientific theories. 
As a consequence, several approaches that focused on mathematical equations rather than logical propositions were developed.
They come in many variants, but generally agree that a theory is constituted by a set of related models. 
While some maintain the goal of axiomatically reconstructing theories \citep[among them][]{suppes60}, others do not, among them \citep{giere88}.
The semantic approach, as it became called, focuses on the content and representation of models rather than merely on their syntactic structure.
Here, the notion of isomorphism has played a strong role in relating the phenomena to the features of the models that describe them \citep{vf80}. 
A theory on this view could be a set of principles, core equations, or axioms from which models can be built, or it could be a set of non-hierarchically related models. These relations can be formal \citep{suppes60} or epistemic \citep{giere88}.
A purely syntactic view of theories will not allow one to describe the turn towards model-independent searches in particle physics described in Sections \ref{sec:data} and \ref{sec:eftintro}, because every QFT, including EFTs, and indeed every Lagrangian, qualifies a model of the theory and thus the distinction in practice that we are hoping to capture would simply be defined away.\footnote{It is true, the quantum field theory used on current particle physics is not axiomatized, but the syntactic understanding of model is often used in a looser sense described here, especially by those who have criticized philosophers' focus on axiomatic quantum field theory.}
In order to assess bottom-up strategies like SM-EFT, we have to look at those philosophical conceptions of models that emphasize their autonomy from the framework theory or even take them as primary and emphasize their epistemic role. 

As a bottom-up strategy, the SM-EFT does not simply extend its empirical content with new fundamental fields, but expands the standard model with higher-order terms consisting of nothing but SM fields.
The goal of such a bottom-up strategy is, initially, not to provide a however limited description of phenomena at a given scale, but to provide a rather broad theoretical framework for experimental searches that may result in constraints on BSM physics.
The philosophical literature on models and theories is vast and we focus on three approaches that are sufficiently different and relevant for our problem.
After sketching the philosophical debate on EFTs, we will present the bottom-up SM-EFT approach in more detail in Section~\ref{sec:classification} and review a case study in Section~\ref{sec:BphysicsConcepts} before applying this discussion to extract philosophical lessons in Section~\ref{sec:analysis}.

\subsection{Models and Theories}
\label{sub:modeltheory}

In moving from syntactic to semantic views, the focus of philosophical analysis shifted from theories to models and this also informed accounts that stressed the autonomy or independence of models from theories, such as the practice-based approaches of \citet{cartwright99}, and \citet{morganmorrison}.
By and large, accounts of models have become broader and more encompassing.
A model no longer needs to be an interpretation of a formal system, a set-theoretic entity, or an isomorphic representation, but can be determined by in part by its function and use in scientific practice.
The distinction between model and theory can be one of degree more than kind, where a theory has broader empirical scope and is more highly confirmed. 
In this Section, we will analyse the distinction in the context of two approaches that have already played a role in discussions of elementary particle physics, the practice-based account of \citet{morganmorrison} and the semantic account of \citet{hartmann1998}. 
While the MaM approach emphasizes the representative and autonomous roles of models in scientific practice, Hartmann's  hierarchy of theories and models attributes a stronger role to framework theories, such as QFT, thus being somewhat closer to Suppes' version of the semantic view \citep{suppes1962}. 
As representation is a key issue, we also introduce the artefactual view of models, given by \citet{knuuttila2011,Knuuttila2017}, that explicitly does not require the accurate representation of models.

On the MaM view, ``the rough and ready distinction followed by scientists is usually to reserve the word model for an account of a process that is less certain or incomplete in important respects. 
Then as the model is able to account for more phenomena and has survived extensive testing it evolves into a theory'' \citep[p.~18]{morganmorrison}. 
\citeauthor{morganmorrison} identify four basic elements of models: 
\begin{enumerate}
	\item They are partially autonomous from both theory and data.
	\item They are purpose-built.
	\item They are representative.
	\item They enable us to learn about the world.
\end{enumerate}
We take it to be fairly uncontroversial that models allow us to learn and that they are built with a variety of purposes in mind.
As such, we will focus our discussion on the remaining two elements.
We will first examine autonomy of models by reviewing the relations between models and theories and move to representation in Section~\ref{sub:reptarget}.
One can see that models are partially autonomous from data and theory from the way they are constructed and function.
Models are not derived entirely from theory or from data, but partly from both and their construction avails itself of scientific practices such as simplification, idealisation, and approximation.
A model is autonomous in that it can be an object of study or a tool in investigations of the world independently of its relation to the formal framework of a theory.

Let us now view the model-theory relation from the perspective of \citet{hartmann1998}, who distinguishes two kinds of theories and two kinds of models:
\begin{itemize}
\item A Type-A theory is a general background theory, quantum field theory in our case. We understand this theory to include a set of techniques like renormalization and only require that it is consistent in a physical sense. 
\item A Type-B theory is a fundamental ``model-theoretical model of a Type A theory, that is a concrete realization of the general formalism of QFT'' \citep[][p.~101]{hartmann1998}. This theory is called fundamental because ``{\it at the moment} everything that can be said about the respective domain of applicability can be captured" (ibid.) by it. 
Hartmann's own examples are QED and QCD.
\end{itemize}
All other theoretical constructs are models.

\begin{itemize}
\item Type-A models are non-fundamental model-theoretical models. They can be simple toy models for a Type-A theory or be as specific as variants or speculative modifications of a Type-B theory. 
Hartmann's example is a quartic interaction $\phi^4$ theory, but it seems to us that also most BSM models belong in this class as long as they do not fall outside the framework of QFT. 
The important point is that these models are models from a formal point of view. 

\item A Type-B model, or phenomenological model, is ``a set of assumptions about some object or system. Some of these assumptions may be \emph{inspired} by a Type-B theory, others may even contradict the relevant theory (in case there is any)'' (ibid.). Accordingly, Type-B models are rather a residual category in Hartmann's fourfold classification. 
They are not necessarily models in a syntactic or semantic sense, but emerge from scientific practice. 
Hartmann's examples are the various models of nuclear and hadronic structure that were developed long before QCD, models which later could be integrated into this formal framework. 
\end{itemize}

Hartmann's understanding of being `fundamental' involves a relationship between the theoretical constructs and the domain of facts known at the time in such a way that if a fundamental model describes a broader domain of facts, it becomes a theory in line with Morrison's above quoted terminology. However, Hartmann seems well aware that his understanding of fundamental is quite permissive and does not support reductionism. Discussing non-renormalizable EFTs as Type B models he concludes that, if one interprets the cut-off physically as their domain of applicability, ``at a given energy scale E a suitably chosen EFT can be fundamental in the sense that (practically) nothing more can be said about the ongoing physics" (p.~104).

Hartmann's examples for Type-B models also correspond to phenomenological models in the MaM approach as examples that ``make use of a variety of theoretical and empirical assumptions'' \citep[][p.~45]{morrison99}.
Their autonomy is based on certain representative features that are not fully derivable from theory or may even go beyond the available background theory, if any.  
Hartmann's requirements for Type-B models seem lower than those of MaM because he only requires `a set of assumptions', and not any partially autonomous representative features. 
Yet his partly hierarchical account emphasizes the role of the background theory and of Type-B theories more strongly, while MaM start from the models' autonomy and their epistemic function. 
In their approach, the existence of what Hartmann calls a Type-A theory is not required---as long there is some framework or some tangible way to study the model. 
Having a well developed framework theory is not excluded in MaM, but the autonomy of models kicks in because the framework theory is insufficient to make clear predictions or to derive the model in a canonical way. This is precisely the situation in particle physics.
\footnote{Note that a hierarchical approach, in the sense of Hartmann and Suppes, also stands behind Karaca's \citeyear{karaca2013} distinction between a framework theory, a model theory (the Type-B theories, in Hartmann's sense) and phemomenological models (Type-A or Type-B models, in Hartmann's sense). \citet{karaca2018} also extends, but qualifies, a similar levelling into models of the detector etc.---which are not of concern here.}

Putting MaM alongside Hartmann's version of a semantic approach we find two, only partially overlapping uses of `model' that one can also discern in the broader philosophical literature. 
One where a model is defined as a restricted or specified part of a scientific theory that is related to a domain of phenomena, and one where a model is built in practice by studying real-world phenomena.  
Both are equally legitimate and are applied in physical practice. 
On some accounts, model building begins within an existing theoretical framework while, on others, it proceeds initially without regard for such a theoretical framework, even though models may well become theories. While MaM clearly belongs into the latter category, Hartmann's account belongs to the former, apart from his introduction of the residual class of phenomenological models.

\subsection{Representation and Target} \label{sub:reptarget}

Let us now turn to the fourth element of MaM, representation.
In contrast to the traditional view where a model must be isomorphic to, or somehow mirror, the phenomenon, representation for MaM is instead taken as ``a kind of rendering---a partial representation that either abstracts from, or translates into another form, the real nature of the system or theory, or one that is capable of embodying only a portion of a system'' \citep[p.~27]{morganmorrison}.
Models can be representative of either the world (their ``target system'') or of theory, they can also represent both, one more than the other. 
For instance, the hydrodynamic model of the boundary layer discussed by \citep{morrison99} demonstrates how representation can happen in both directions.
Prandtl built a small water tunnel that acted as a representation of the world, which exhibited different behaviours in different regions and allowed him to construct a mathematical model.
The mathematical model in turn represented aspects of both the classical theory and the Navier-Stokes equations, neither of which could be applied directly to the real world. 
Thus, the models represented in different ways, had different targets of representation, and acted as mediators between theories and the world. 
Morrison contrasts the boundary layer model, where one has to develop a target for mathematically efficient representation, to wit, the boundary layer, with the mathematical pendulum, where the formalism of Newtonian mechanics provides all the formal tools for a de-idealisation, i.e., possible terms for the introduction of friction, finite size of the bob, etc. 
It is this emphasis on the practical aspects of representation that distinguishes MaM from Hartmann's more theory-based approach---even though Hartmann's residual category (Type-B models) is pretty elastic.

In our use of representation, we follow MaM and make no strong demands on isomorphism or accurate representation and gladly accept the important role of modelling practices, such as idealisation, approximation, and abstraction. 
Exactly how the elements of a system are mathematically represented in a given model can only be justified by some story made in a case-by-case basis \citep[cf.~]{hartmann1999}. 
We are thus willing to endorse a rather liberal notion of representation where a representation in a scientific model is a mathematical description of a target or target system. 
A target (or target system) can be whatever the model is aiming at describing, which may be actual, as in the case of beta decay, or merely potential, as in models of BSM physics. 
Targets may be objects, such as fields, or descriptions of how objects interact, such as interaction vertices, and target systems may include both. 
Targets may be as accurate as possible, or they may be idealised, abstracted, and approximate. 
Again, we make no strong demands on the `mapping' relation. However, a bare minimum condition for being a model is that it must be a \textit{model of something}, and as such it must at least have some target of representation. 
Furthermore, its target, whatever it is, needs to be distinct and well defined, such that we can meaningfully say that it is being modelled.
Having a specifiable target is then a bare necessary condition for modelhood, though it is of course not a sufficient condition.
In our specific context, these targets are elements are of physics beyond the SM, such as new fields and vertices.
For a genuine model of BSM physics, its target is some purported phenomenon beyond the SM and what it represents needs to be distinct from the SM. 
If a BSM model has a distinct and well-defined representational target it will allow physicists to make specific predictions about new physics.
Making numerical predictions alone, however, is not sufficient for having a well-defined target. 
Lagrangians like the one in Eq.~\ref{eq:smeft} contain a large number of parameters that can be tweaked to absorb any deviation. 
The essential point of Eqs.~\ref{eq:BSM} and \ref{eq:smeft} is thus the schematic separation into a well-defined background model---the SM whose sizeable number of parameters are considered fixed by experiment---and deviations from it that are parametrized in terms of an EFT.

While the MaM approach requires that a model has some target or even representative features, others have argued  
that representation is not a characteristic feature of a scientific model.
Some practice-based approaches consider representation an accomplishment of model-users rather than a feature of the model itself.
\citet{knuuttila2011,Knuuttila2017} and \citet{teller01} have argued that there is no general account of how and in virtue of what scientific models can be said to be adequate representations. 
In a sense, anything can stand for anything and what represents something is what is chosen to represent it.
This representation can be successful or not for a given end, but whether it represents or not is not decided by features of the model or the target.
On accounts such as Knuuttila's, the key features of models are epistemic: they are tools that allow us to learn, to make good inferences, to convey useful information about the world, and the role of representation is no longer central.
Knuuttila argues that representationalist views of modelling come up against problems because they treat models as distinct objects that need to be mapped to the world in order to justify their success. 
By contrast, she develops an artefactual account that attempts to dissolve these problems by conceiving of models as epistemic tools rather than abstract objects.

This ``amounts to regarding [models] as concrete artefacts that are built by specific representational means and are constrained by their design in such a way that they facilitate the study of certain scientific questions, and learning from them by means of construction and manipulation'' \citep[p.~262]{knuuttila2011}.
She lists five characteristics of models as epistemic tools:
``(i) the constrained design of models, (ii) non-transparency of the representational means by which they are constructed, (iii) their results-orientedness, (iv) their concrete manipulability and (v) the way their justification is distributed so as to cover both the construction and the use of models'' \citep[p.~267]{knuuttila2011}.
The epistemic characteristic of models does not include accurate representation, but neither is it precluded.
Representation is always an accomplishment. 
But a model's success cannot be assessed without ``the culturally established external representational tools that enable, embody and extend scientific imagination and reasoning'' \citep[p.~1]{Knuuttila2017}. 
They ``form an ineliminable part of the model itself,'' (ibid., p. 3) not just of its description. 
``What distinguishes models from many other kinds of representations is often their systemic, autonomous nature,'' (ibid., p. 17) which they share with fictions.
Knuuttila illustrates her artefactual account with the example of highly-idealised mathematical models that are applicable in unrelated empirical contexts.

\subsection{Effective Field Theories}\label{PhilEFT}

There already exists considerable philosophical literature on effective field theories.
Philosophical attention turned to EFTs in the late 1980s and early 1990s with the seminal papers by \citet{teller89}, \citet{cao1993}, and \citet{huggetweingard}. 
\citet{cao1993} used EFTs to articulate anti-foundationalist metaphysics where there is no final theory, but only an infinite tower of EFTs. 
From these discussions various metaphysical and ontological arguments were picked up in \citet{castellani2002} and \citet{fraser2009}, and more recently in \citep{bain2013a,Crowther2016-CROESU,Rivat2020-RIVPFO,Ruetsche2018,williams2018}. 
\citet{Rivat2020-RIVPFO} provide an excellent overview of the debates about EFTs in philosophy and note that they are primarily about issues of ontology, emergence and inter-theory relations, and fundamentality. 
A substantial part of these debates focuses on whether they are robust or fundamental enough to be worthy of philosophers' attention as compared to axiomatized approaches---which, for \citet{wallace2006} and \citet{williams2018} are still in the focus of philosophical orthodoxy. 
Williams' `effective realism' about EFT is certainly one way to justify the value of EFTs through the robustness of entities, properties, or relationships that are "accessible (detectable, measurable, derivable, definable, producible,
or the like) in a variety of independent ways," \citep[p.~219]{williams2018}.  
\citet{fraserj2020} defends a more refined selective account of realism based on  the renormalization group. 
A detailed criticism of the various ways to identify the real parts of EFTs has been given by \citep{rivat2020a}.
Another, more instrumentalist, view was developed, initially by \citet{huggetweingard}, in response to the controversial views of \citet{cao1993}, whereby EFTs are merely tools for extending the scope of physics. 
\citet{Rivat2020-RIVPFO} describe Huggett and Weingard as saying that 
``one should not draw conclusions about the hierarchical structure of the world, the existence of a fundamental level, or the prospects of a completely unified theoretical description'' based on EFTs.
This pragmatic view is picked up on by \citet{hartmann2001}, for whom EFTs do not suggest anti-foundationalist metaphysics, nor will they eclipse models or theories, but rather that they have a role play in the search for new physics alongside models and theories.
According to \citet{Butterfield_2010}, the use of EFTs come from ``an opportunistic or instrumentalist attitude to being unconfident'' about the applicability of QFT at high energies.
\citet{Grinbaum2008-GRIOTE} also endorses this kind of view and likens the current interest in EFTs to the widespread use of S-matrix theory in the 1950s, where a clear consensus on the way forward was missing. 
With considerable foresight, Grinbaum suggests that ``perhaps something like this is happening today with EFT and the model-independent analysis of new physics'' (ibid., p.~45).
Indeed, today, physicists are unsure what the next accepted theory will be like and once again we find them turning to pragmatic benefits in the form of EFTs.

But we have to note that most of these authors have focused on EFTs in general and not specifically on the bottom-up approach of the SM-EFT.
\citet[p.~3]{Rivat2020-RIVPFO} even allude to SM-EFT but consider it as one among many black-box models. 
\citet[p.~72]{Crowther2016-CROESU} considers bottom-up EFTs as the main tool for constructing a tower of EFTs in the style of Cao and Schweber, which she does not consider ``as a threat to the possibility of there being an ultimate fundamental theory of physics'' (p.~74).  
One exception is \citet{wells2011} who examines bottom-up EFTs as a part of a general method, or mind-frame, for scientific research that could have been used, he argues, to point the way towards general relativity by predictions about the perihelion precession of Mercury already in the late 19th century. 
The in-principle feasibility of such a bottom-up approach, to his mind, suggests that physicists should focus on improving theories through EFTs and not be exclusively concerned with principles and justifications.
Wells does not discuss whether this makes EFTs or effective theories in general similar to models.

We will largely avoid issues of realism or reductionism because physicists availing themselves of SM-EFT are openly granting the reality of the SM in order to look for deviations of any kind.
Their eventual goal is to find constraints on possible models of such new physics. 
We are interested in the use of EFTs in practice and the importance of the bottom-up approach and find the modelling perspective to be more illuminating.
Thus we take our discussion to pick up on questions introduced by \citet{hartmann2001}, who takes a detailed look at the history and development of EFTs in the context of models and theories.
The metaphysical and ontological theorizing based on EFTs, which Hartmann was reluctant to do, has been strongly taken up in the last two decades, but his discussion of the roles of EFTs, models, and theories has been more quietly received.
In looking at case studies in hadron physics, he argues that EFTs produce scientific understanding of the processes they are modelling; they are simple, intuitive, and satisfy a variety of pragmatic and cognitive goals. 
There is rather an interplay between theories, models, and EFTs in physics research and distinguishing them is not always easy. 
``EFTs share many of the functions of theories and models. Like models, they provide a local, intuitive account of a given phenomenon in terms of the degrees of freedom which are relevant at the energy scale under consideration. They are relatively easy to solve and to apply, and they are heuristically useful\ldots Like theories, EFTs are part of a bigger picture or framework, from which they can be derived in a controlled way. They help to make predictions and to test the theory they relate to'' (\citeyear[p.~294]{hartmann2001}). 
Among Hartmann's examples are the above-discussed Fermi theory and the V--A theory of the weak interaction. 
Our analysis similarly finds that the SM-EFT has complex relations to theories and models. 
While accordingly this double relationship remains, there are significant differences and we will argue that its nature may even depend on the experimental success of the SM-EFT.


\section{SM-EFT: The Bottom-Up Approach}
\label{sec:classification}

Currently, in particle physics, there are no experimental
results that would allow one to clearly identify any of the 2499 additional possible
dimension-six operators
(see Sec.~\ref{sec:eftintro}) as a signal of BSM physics. 
Instead, only more or less strong upper limits exist.
As we will discuss, the way the SM-EFT is used by physicists can be divided into three different stages.
These stages are used to support the search for BSM both experimentally and theoretically,
with several goals coming together: they serve as accounting scheme both in general and in an approach focussed on a special sector of the SM. 
Thus, they impose constraints on model building and, in case of an indication for a non-vanishing
Wilson coefficient, allow physicists to evaluate potential general consequences and possible 
realizations in a concrete BSM model.

We refer to the SM-EFT with all 2499 dim-6 operators as \emph{Stage 1}.
This stage 1 SM-EFT does not represent a specific new interaction
and thus it has no target in a sense that would allow it to
guide experimental searches.  
If all coefficients $c_i$ in Eq.~\ref{eq:smeft}
were considered to be potentially non-zero, it would allow a virtually infinite number 
of observable distributions (be it interaction rates, asymmetries, kinematic
distributions or any other type of possible particle physics measurement) 
to deviate from the SM without
any specific theoretical preference of what should be experimentally
expected.
This is not sufficient to constitute an experimental target. 
Thus, the stage 1 SM-EFT just serves as a tool, an accounting scheme, to potentially parametrize, or in general
constrain, possible deviations from the SM. 

\emph{Stage 2} of the bottom-up approach is an operational response
to the impossibility to theoretically control 2499 parameters or the corresponding experimental distributions simultaneously. 
Instead, physicists typically just focus on a small subset of the SM-EFT and attempt to include all relevant experimental information from one specific experimental sector, prominently undertaken for top quark physics, bottom quark physics (see Section~\ref{sec:Bphysics}), or Higgs physics. 
Such separations between different sectors of the SM
violate gauge invariance, but a reasonable choice of the basis of the
SM-EFT (see Section~\ref{sec:smeftphysics}) alleviates its effect on  
finding deviations at the LHC in the respective sector.
Since currently the strongest focus is on Higgs physics, as a new
and unique potential portal for new physics, let us clarify the stage 2 concept with
this example~\citep{Espinosa:2016ovf,deFlorian:2016spz}.
The number of operators for such dedicated studies are given by a suitable choice
of the basis, their relevance to Higgs couplings,
experimental constraints on the corresponding Wilson coefficients and
possibly by assuming additional symmetries.
While for such a `Higgs EFT' different approaches exist, they typically neglect
four-fermion operators, operators
describing gauge-boson self-interactions, operators for the $W$-mass and 
for the fermion couplings to $Z$ and $W$.
This reduces the number of independent operators to 59, which still far exceeds the experimental capability to set constraints simultaneously on the parameters from separate experimental sectors. 
However, measurements already put significant constraints on several of these parameters, such that for experimental and theoretical studies, only part are considered relevant for LHC physics.
In addition, the number of operators can be reduced to a mere handful by assuming CP conservation or minimal flavour violation.
This reduction to a few parameters is operationally motivated and makes \emph{Stage 2} experimentally and phenomenologically useful by mapping measurements to a manageable set of parameters. 
But even while the number of operators potentially deviating from the SM is reduced, this sectorised SM-EFT still lacks a specific experimental target. 
It only reduces the number of distributions that appear to be a meaningful probe for deviations.
This conclusion is also supported by the current experimental constraints on the Wilson coefficients in the Higgs boson sector: for a model with a clear experimental target, one would expect that experimental results allow one to constrain the allowed range of parameters or to refute the
model. 
However, as shown in~\citep{ATL-PHYS-PUB-2019-042},
the SM-EFT parameters cannot be constrained if more than two Wilson
coefficients are tested at the same time: there are ``flat
directions'' for all parameters, which means that effectively every
parameter can take any value. Therefore, it carries no
\emph{specific} information based on the data.

An EFT analysis obtains a clear target if specific operators are selected, either to account for experimental indications of a new signal, or by specific additional theoretical motivations reaching beyond the logic of the SM-EFT. 
We refer to it as \emph{Stage 3}.  
In such a situation, only a very small set of Wilson coefficients would be considered to be non-zero, such that only few operators are to be considered relevant.
An example of such a situation is given in Section~\ref{sec:BphysicsConcepts}.
In this case, it would be possible to meaningfully reduce the
ambiguities in representation stemming from the different bases. 
The most efficient SM-EFT basis that describes the data with the
smallest set of operators would then represent a small subset of the
possible new effective vertices.  
This is not dissimilar to the
development of the Fermi theory (see~\ref{ssec:Fermi}). 
In the Fermi
theory, one 4-point interaction vertex with one associated coefficient
$c^2=G_F$ is added to the previously known physics in an effective
theory. 
In case of the bottom quark analyses, described in Section~\ref{sec:BphysicsConcepts}, the apparent violation of lepton-flavour universality in the bottom quark sector motivates the use of specific operators to describe observed experimental features, thus they represent specific interactions. 
This is a qualitative difference in Stage 3 with respect to Stages 1 and 2: the SM-EFT no longer describes potential non-resonant deviations from the SM in a non-unique way (as in Stage~2), but in Stage~3, one selected basis of the SM-EFT uniquely describes features that are distinctively different from the SM.

\subsection{Differences between these Stages}

Differences between these stages highlight issues that will become
relevant for our discussion of the model character of SM-EFT.
Let us first remark that these stages are meant as analytical categories and do not have to be marched through sequentially in actual scientific practice.
For example, in the early 1980s, without reference to
a systematic development of the SM-EFT,~\cite{Eichten:1983hw} proposed
a model for fermion substructure (`preons') in analogy to the Fermi theory.

The major difference between Stages 1 and partly Stage 2 versus Stage 3 is the (absence) of an underlying specific physics problem:
the SM-EFT of Stage 1 is motivated only by the general and unspecific idea that the SM should be embedded in a larger theoretical framework and the restrictions of Stage 2 are largely motivated by considerations of scientific practice, Stage 3 is built either to understand a measurement beyond the SM, as will be detailed for a current example in Sect.~\ref{sec:BphysicsConcepts}, or by a concrete idea of how BSM should look like, be it the preon-hypothesis of~\cite{Eichten:1983hw} or the violation of baryon number in the framework of Grand Unified Theories (GUTs), as, for example, in~\cite{Weinberg:1979sa}.   
As already discussed in Sect.~\ref{sec:physrepres},
SM-EFT at Stage 1 works as an accounting scheme to quantify deviations
from SM distributions and possibly correlate different operators.
As such, the SM-EFT is a convenient tool to globally assess the status of the SM.
In contrast, selecting one or a few operators in Stage 3 is a step towards the goal of physicists to arrive at a concrete BSM model involving new entities.
The roles of target, prediction, testability, and representation are therefore significantly different in these stages, which can be gleaned, for example, from the paper of~\cite{Weinberg:1979sa}. 

In his paper, Weinberg addresses the specific problem of B--L (non-)conservation (B denoting the baryon number, L the lepton number) with GUTs in mind.
Thus, he explicitly neglects the other operators of the SM-EFT because they are of no relevance for his specific target, but he is careful to select dimension-6 operators that do not conserve baryon number. 
From those he derives general constraints for B--L violation, including relations among observables like partial decay widths of baryons.
Weinberg thus starts from Stage 2; from a well-defined theoretical target, which implies an experimental target, viz. proton decay.
Both of these were considered in concrete models before his paper.

On the other hand, Weinberg also makes clear that limitations exist in his EFT approach. 
He immediately emphasizes that B--L non-conservation should be understood better by specific gauge models of baryon decay \citep[p.~1569]{Weinberg:1979sa}.
A derivation of such a model is hardly possible if it should simultaneously address all 2499 operators of the SM-EFT.
Furthermore, Weinberg's paper is sometimes used as an argument for the ability of SM-EFT to predict: in focussing on two specific dimension-5 operators, Weinberg states that these would produce a neutrino mass between roughly 10$^{-5}$ and 10$^{-1}$ eV.
Here, Weinberg argues from what we call Stage 3, which we agree, has a target and representational content.
This, however does not mean that an EFT \textit{per se} allows one to predict: to actually arrive at the values for neutrino masses,
Weinberg assumes certain Wilson coefficients, a cut-off scale of 
10$^{14}$ GeV, suggested by the merging of the electromagnetic, weak and strong couplings, and a coupling of $\cal{O}$(1).
Thus, the EFT \textit{simpliciter} points out at most possibilities. 
To provide a quantitative prediction, it has to be supplemented by an additional assumption that comes closer to a concrete model.
The same is true for the preon assumption of~\cite{Eichten:1983hw}: it motivates searches in certain processes, which were anyhow motivated with parametrisations other than EFT, it provides a framework to set constraints on fermion substructure, but, in the absence of a quantitative estimate of observable effects, it does not allow one to definitely rule out such substructure. 

The inability of an EFT at any stage to predict without further model assumptions implies that it cannot be meaningfully tested.
While one can certainly look for any kind of deviation the SM-EFT allows, something that is anyhow in the spirit of expecting BSM somewhere, any non-observation of a deviation would not falsify SM-EFT since it does not allow for a quantitative prediction.  
In consequence, the SM-EFT cannot be tested in a meaningful sense; 
a single operator, supplemented by a physics model, can. 

We thus diagnose a major distinction between Stages 1 and 2 and Stage 3 that will be relevant for determining whether and at what stage the EFT is a model. 
At first glance it may be surprising that the SM-EFT, if interpreted as a collection of operators, does not produce a target or have representational character, whereas each individual operator has.
However, we understand this as a breaking of an (virtually) infinite amount of possibilities into just a single one.
As we know from many examples, such a breaking significantly changes the character of the related objects.
  


\section{Exemplifying Search Concepts in Bottom Quark Physics}\label{sec:BphysicsConcepts}

In Section~\ref{sec:classification}, we discussed a classification
scheme for the SM-EFT, where the experimental status of the theory can
play an important role in the reduction of very many degrees of
freedom to a very small number. The SM-EFT in \emph{Stage~3} thus can
gain representative character through the identification of new
vertices beyond the SM, which represent new types of interactions
between SM particles. One such experimental situation, which has been
emerging over the last years, are precision tests of rare leptonic
decays of $B$ mesons, where lepton universality is
violated in experiments
at a statistical significance of around ${\cal O}(3\,\sigma)$ on
average for sensitive single measurements \citep{Aaij:2015oid,Aaij:2017vbb,Aaij:2019wad,Abdesselam:2019dgh,Abdesselam:2019wac,Buttazzo:2017ixm}.
In this section, it is shown how the different BSM approaches are applied to this heavily discussed set of physics processes.

In the SM, the electroweak gauge bosons, $Z$ and $W^\pm$, couple to
the three lepton flavours ($e$, $\mu$, $\tau$) universally (i.e.\ with
the same strength).  Recent measurements of B-meson decays such as
$b \to sl^+l^-$ and $b \to c l \nu$ show hints for the violation of
lepton flavour universality (LFV) (for an overview, see
also~\citep{Albrecht:2018frt}).  As of yet the statistical and systematic
uncertainties are too high to allow firm claims, but if confirmed,
these measurements would provide evidence for BSM physics. To evaluate
LFV, the three approaches introduced in
Section~\ref{sec:eftintro} to evaluate LFV are pursued: concrete
models; Stage~2 and Stage~3 EFTs; and
simplified models. 

\subsection{Concrete BSM Models}
\label{sec:B-con-BSM}

To explain the apparent LFV with a concrete model, new particles with
different couplings to leptons are assumed.  Furthermore, since LFV
has only been observed for the bottom quark, it is suggestive to
assume such a particle to have an affinity to bottom quarks, or in
general to quarks of the third generation.  Attempts to explain the B
anomaly include SUSY~\citep{Altmannshofer:2017poe}, strongly
coupled models, such as composite Higgs~\citep{Greljo:2015mma}, or
additional (heavy) Z's or W's~\citep{Boucenna:2016qad}.

While these models may be easily adjustable to work for LFV in 
bottom hadron decays, they also have implications for other processes
that cannot be accommodated easily and require a high level of fine
tuning. Furthermore, some of these hypotheses, although starting from
a different underlying concept, arrive at a similar phenomenology for 
LFV effects, and thus cannot be experimentally distinguished from each other. Therefore, physicists currently turn to EFTs and simplified models, which
provide general constraints on a viable model, 
despite the fact that concrete models would have been more satisfying.
This is why several authors prefer a bottom-up approach using the
SM-EFT (see e.g.~\citet{Buttazzo:2017ixm}).

\subsection{B-physics Anomalies: Effective Field Theory Approach}\label{sec:Bphysics}

The physics of bottom hadrons involves different scales, the QCD scale of
some 100 MeV, the scale of the bottom quark of some 4 GeV and the
electroweak scale of some 100 GeV. It is thus adequate to use an EFT to
describe bottom quark decays.  The corresponding Lagrangian, containing only
transitions of the bottom quark into other SM particles, is a subset
of the SM-EFT and has been used in analyses of bottom decays and
transitions for some 20 years.

While the indication of LFV can be analysed with a full EFT for bottom
quarks, the observed properties indicate only certain operators to be
relevant as pointed out in~\citet{Buttazzo:2017ixm}. Conveniently the
(potential) BSM effects can be represented as an additional
contribution to the SM Lagrangian. For example in
\citet{Buttazzo:2017ixm} the effective Lagrangian for
$b\rightarrow s l^+l^-$
\begin{equation}\label{eq:leff}
{\cal L}_{\rm eff} = {\cal L}_{\rm SM} - \frac{1}{v^2}\lambda^q_{ij}\lambda^l_{\alpha\beta}\left[C_T(\overline{s}^i_L\gamma_\mu\sigma^a b^j_L)(\overline{e}_L^\alpha\gamma^\mu\sigma^a e_L^\beta)+C_S(\overline{s}_L^i\gamma_\mu b_L^j)(\overline{e}_L^\alpha\gamma^\mu e_L^\beta)\right]
\end{equation}
is postulated for the production of $e^+,~e^-$ (or $\mu^+ \mu^-$ by
replacing $e_L$ by $\mu_L$).\footnote{Instead of general quark
  transitions, here only the one relevant for the indication of LFV,
  namely the $b\rightarrow s$ transition is used} It contains two
four-fermion operators which describe new interactions between
left-chiral quark and lepton fields $(s,b)_L$ and $(e,\mu )_L$,
respectively, already basically known from Fermi theory.  The
coefficients $C_T$ and $C_S$ are flavour blind and encode the new
physics scale $C_{T,S} \propto \ v^2/\Lambda^2$, where $v \approx
246$\,GeV is the vacuum expectation value of the SM Higgs field. The indices $T$ and $S$ refer
to the colour-triplet and colour-singlet structure of the
corresponding operators, respectively.  The flavour structure of the
new interactions is determined by a $U(2)_q\times U(2)_l$ flavour
symmetry and encoded in the Hermitian matrices $\lambda^q_{ij}$ and
$\lambda^l_{\alpha\beta}$. The $C_i$ and $\lambda $ coefficients are
free parameters.

\begin{figure}[tbh]
\begin{center}
      \includegraphics[width=0.75\textwidth]{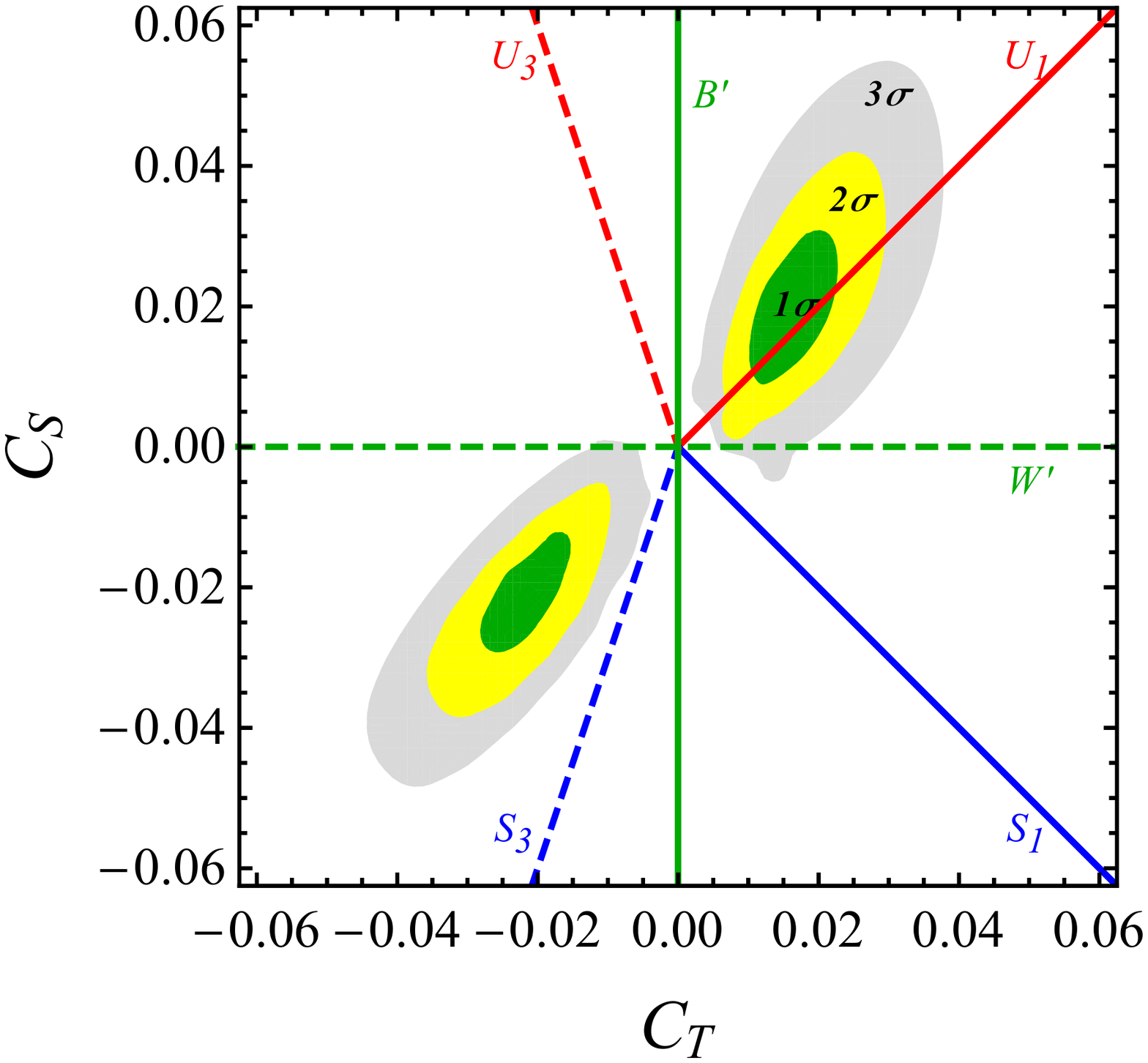} 
      \vspace*{-30mm}
   \caption{Fit of the coefficients $C_T$ and $C_S$ in Eq. \ref{eq:leff} from flavour physics observables. 
   	The parameters $\lambda^q_{sb}$ and  $\lambda^l_{\mu\mu}$ have been marginalised over. 
   	The green, yellow, and grey regions show the 1, 2, and 3$\sigma$ contours, respectively. 
   	Also shown as the straight lines are the correlations between $C_T$ and $C_S$, as predicted in various simplified models with a single mediator: colour-less vectors are shown in green, coloured scalars in blue, and coloured vectors in red. 
   	Electroweak singlet mediators are shown as solid lines, triplets as dashed lines \citep[from][]{Buttazzo:2017ixm}.}
\label{fig:fit}
\end{center}
\end{figure}

Describing the BSM contribution in the EFT approach allows one to
combine the apparent flavour violation in $b \to sl^+l^-$ with other
measurements both in the bottom quark sector but also in LHC production of
pairs of leptons of different flavour. These can be used
to further constrain the free parameters.  A global analysis of
existing flavour data \citep[see, e.g.,][]{Albrecht:2018frt} indicates
non-zero coefficients $C_T$ and $C_S$, as shown as the green, yellow
and grey regions (1, 2, and 3$\sigma$ contours, respectively) in
Fig.~\ref{fig:fit} \citep{Buttazzo:2017ixm}.  This is an example of
how the SM-EFT acquires representative character: Committing to the
existence of non-zero values of $C_T$ and $C_S$ corresponds, in the
chosen basis of the EFT, to the existence of two new vertices of SM
fields, namely two different types of a direct coupling between four
different fermions which is not present in the SM: two different
quark flavours and two leptons per vertex. Assuming the existence of
these two new vertices is analogous to the assumption of the existence
of the Fermi interaction before the SM was conceived. Also, these
vertices can then be specifically studied for their effects on
precision measurements in other experiments, or e.g.{} for their
predictions of distortions of kinematic spectra away from the SM in
LHC measurements involving the same quarks and leptons as the
B-physics measurements.

\subsection{Simplified Models}
\label{sec:B-SimMod}

The allowed regions in $C_T$ and $C_S$ from the EFT results are not
transparent in terms of an underlying physics idea. However,
they can both be
interpreted as new interactions in a \emph{Stage 3} SM-EFT and can be related
to elements of the concrete BSM models of Sect.~\ref{sec:B-con-BSM} in
a simplified model approach.  As discussed in Sect.~\ref{sec:SimpMod},
the idea of simplified models is to focus on just part of a BSM model, say a
single type of particle and evaluate its constraints and potential
consequences. This approach starts from a Lagrangian consisting of the
SM and a term representing a hypothetical new particle with a free
coupling strength.  The outcome of this Lagrangian is then matched to
Eq.~\ref{eq:leff}.
Assuming $B'(1,1,0)$ or $W'(1,3,0)$; (colour-triplet scalar particles)
$S_1(\bar{3},1,1/3)$ or $S_3(\bar{3},3,1/3)$; (colour-triplet vector
particles), $U_1^\mu(3,1,2/3)$ or $U_3^\mu(3,3,2/3)$\footnote{The
  numbers in brackets indicate the colour, weak, and hypercharge
  quantum numbers.}, one obtains a certain correlation between $C_T$
and $C_S$, shown as straight lines in Fig.~\ref{fig:fit}.  While in
these simplified forms the particles are only defined by their quantum
numbers, they could be interpreted as new electroweak bosons, scalar
or vector leptoquarks, which already were discussed in the context of
concrete BSM models.  Comparing the analysis performed within the
EFT, and the predictions of the different
simplified models, the model with a coloured vector mediator,
$U_1^\mu(3,1,2/3)$, is clearly preferred.

\section{Philosophical Lessons}	
\label{sec:analysis}

In this final section, we develop four lessons from the foregoing discussion and investigate their consequences for the present philosophical debates on models. 
First, we find that the SM-EFT cannot be categorized as simply as an explicit BSM model or as a simplified model. 
We demonstrate this by looking at the MaM approach (\ref{sub:autonomy}) and Hartmann's version of a semantic (\ref{sub:modelhood}) approaches. 
Second, even though the three stages presented in the previous section must be considered as parts of a coherent epistemic strategy, the autonomy and representational features are different for the stages, affecting their respective modelhood (\ref{sub:smeftrep}).
Third, we argue that the fact that one applies the same theoretical and experimental techniques in the SM-EFT approach as are used in the study of BSM models and simplified models, does not automatically render the SM-EFT a model (\ref{sub:epistemic}). 
This shows, to our mind, why, more generally, it might be attractive to investigate the boundary of the model concept rather than advocate a liberal use. 
Fourth, we argue that focusing on models and their representative features can provide an ontologically lean perspective on EFTs insofar as it does not require to take a general stance on realism, in contrast to large part of the present debates (\ref{sub:EFT}).

\subsection{Is SM-EFT an Autonomous Mediator?} 
\label{sub:autonomy}

As noted in Section~\ref{sec:models} above, it is rather uncontroversial that models are constructed to fulfil specific purposes and that they help us learn about a theory or a target system, respectively.
These basic elements of the MaM approach not only apply to BSM models and simplified models, but also to SM-EFT, whose purpose is efficiently accounting for possible deviations in order to possibly place constraints on models of BSM physics. 
The partial autonomy from data and theory, another element of MaM, makes for a more interesting analysis.\footnote{We will discuss the fourth element, representation, in Sec.~\ref{sub:smeftrep}.}
Let us start with the autonomy of explicit BSM models.
In \citep{maettigstoeltznersig} it was argued that BSM models stand in a twofold mediating relationship because they are placed within the framework of quantum field theory or a more general successor theory and they must simultaneously recover the SM as a low-energy limit. 
Or put otherwise, they must recover the SM as a top-down EFT.
From this, we see that SM-EFT is at least partly autonomous from data. While it features all the SM fields whose properties must accord with data, the many additional terms are only determined by general theoretical considerations, such as the SM symmetries.

\citet{mccoymassimi} have argued that simplified models are autonomous from data and `theory' (here meaning full BSM models, like supersymmetry) because a simplified model is partly independent from any particular BSM model. 
Due to their autonomy, they argue, simplified models can play a role mediating between data and BSM models analogously to the mediating role played by models in the original MaM account. 
Yet, this autonomy must be qualified because the particle content of a simplified model is borrowed from a potentially large class of BSM models. 
For example, a simplified model may feature a new scalar top-like particle, but such a particle has originally been proposed within a class of full BSM models. 
The main point is that simplified models do not specify all properties of the additional fields, admitting a large space of physical possibilities.
McCoy \& Massimi call such modelling perspectival because it does not require representation in the sense of a mapping onto a target or ``onto an actual worldly-state-of-affairs (or suitable parts thereof) but representation has instead a modal aspect: it is about exploring and ruling out the space of possibilities in domains that are still very much open-ended for scientific discovery" \citep[p.~338]{mccoymassimi}.
The initial space of possibilities in the SM-EFT is much larger than the one of simplified models because, at least initially, it is no longer motivated by any properties of classes of models, outside of the SM. 
At Stage 1, there is no clear and distinct perspective to take. 
Thus, the perspectival modelling cannot be further extended to consider SM-EFT a model.

SM-EFT is not autonomous from theory in that it uses a theoretically well-established procedure in expanding the Lagrangian into operators of higher dimensions. 
Only fields of the SM theory are used and the basic symmetries of the SM are retained. 
Hence the SM-EFT is, at Stage 1, determined by general theoretical principles and a well-confirmed theory, the SM, but not by any particular BSM models; all formally possible terms are taken into consideration.  
Here, one operates in a generic space of possibilities that parametrizes our ignorance and can be used for exploratory experimental searches. 
According to \cite{wells2012}, this generic approach is even a requirement of mathematical consistency, to avoid divergences and maintain the independence of arbitrary scale cut-offs. 
Leaving out terms destroys the renormalizability of the theory, and if one picks out certain terms the corrections might be associated with terms that had been left out. 
Thus at Stage 1, SM-EFT has little autonomy from the SM and the framework theory. 
In Stage 2, any such autonomy is only apparent because it arises as the product of pragmatic decisions for the sake of tractability or simplicity.
But even if the mathematical consistency is lost, a Stage 2 EFT for special sectors is largely determined by the SM in its field contents and symmetries. 

A Stage-3 EFT is, however, sufficiently autonomous from theory that it could be considered a model. 
This is not only the case in terms of MaM's notion of autonomy. 
Also \citet{Crowther2016-CROESU}, more specifically, argues 
that ``the idea of autonomy (or, rather, quasi-autonomy) in EFT comes from the fact that the low-energy theory is largely independent of the details of the high-energy theory. There is extra information contained in the high-energy theory, far over-and-above that required in order to describe the low-energy behaviour of significance.'' 
This autonomy makes the EFT robust against changes in the microphysics and together with the fact that is a novel description, implies for her that EFTs are emergent.  
A Stage-3 EFT can simply be written down to model some new interaction, in which case, its autonomy from the full theory (as well as a clear target of representation, and hence modelhood) is introduced by hand. 
However, one may also arrive at a Stage-3 EFT based on the systematic expansion in SM-EFT, where the relevant EFT operators are selected by deviations from the SM observed in measurements. 
In this sense, SM-EFT is an example of a potentially successful bottom-up strategy that accomplishes autonomy from the theoretical description in terms of products of SM-operators. 
In such a case, the SM-EFT approach would become the proverbial Wittgensteinian ladder; a formal theory to be thrown away and replaced by a concrete BSM model introducing a new field, a simplified model, or a vertex model of some new phenomenon in the same style as the Fermi theory had done for beta-decay.

\subsection{SM-EFT in Hierarchical Approaches}
\label{sub:modelhood}

Let us now return to the problem handed to us by \citet{hartmann2001}, namely the difficulty in disentangling EFTs from models and theories and address it in the context of Hartmann's (\citeyear{hartmann95}) hierarchical approach discussed in Section~\ref{sub:modeltheory}. 
Theory here has to be understood in the double sense of a Type-A theory (such as QFT) and a Type-B theory (such as the SM). 
All explicit BSM models can be subsumed under Hartmann's classification, although it is sometimes a matter of how `fundamental' their aspirations are and whether there are presently any real or purported phenomena they describe.\footnote{Recall that Hartmann's notion of `fundamental' is defined with respect to the present experimental knowledge and does not involve any reductionist aspirations.}
Supersymmetry (SUSY) e.g. may be considered a Type-B theory 
that captures all that can be said in its intended domain of applicability; it is true, key features of SUSY have not been empirically validated. 
However, it can also be interpreted as a Type-A model if one interprets SUSY just as one way to provide inputs for a more fundamental theory. 
This can be motivated by the different symmetry breaking mechanisms invoked in special realizations of SUSY. 
Other concrete BSM models, like technicolor and Randall-Sundrum models, do not have the rather fully worked out fundamental structure, such that they would be considered as capturing their total domain of applicability. 
In this sense, they can be classified as Type-A models in virtue of their serving as `input for more fundamental constructs'. 
However, there may also be an emphasis on the phenomenological aspects of these models in the sense of Type-B models introducing a `structure that cannot be deduced from that theory'. 
Such a characterization is certainly appropriate if models (partly) work outside the framework of QFT; variants of extra-dimensional modes can be counted as such. 
Given that simplified models are developed from BSM models, they should be classified as Type-A or Type-B models depending on how their space of possibilities is configured.
Recall that, as mentioned in Section~\ref{sub:modeltheory}, Hartmann's requirements for a Type-B model are lower than in the MaM approach; thus it seems to us plausible that the reasoning of McCoy and Massimi concerning representation can be copied to obtain a perspectivalist generalization of Hartmann's approach.

Since for Hartmann the Fermi theory is an EFT and a phenomenological model, one may wonder whether SM-EFT overall can be considered as a model in a non-trivial sense, that is, beyond the fact that any SM-EFT is formulated in terms of a Lagrangian and thus a formal model of QFT.
As argued in Section~\ref{sub:autonomy}, SM-EFT is, at Stage 1, primarily influenced by theoretical considerations, such as symmetries and renormalization. 
It is a formal technique allowed within the context of a Type-A framework theory and strongly dependent on the existence of a well-established and well-tested model, or a Type B theory, such that the formal operators of higher order can be experimentally analysed as deviations from the SM. 

But in contrast to concrete BSM models or simplified models, SM-EFT at Stage 1 is neither a toy model \citep[see][]{Reutlinger2018} that could provide understanding of a Type-A or Type-B theory, nor a variant of a Type-A or Type-B theory. 
By integrating all formal possibilities, SM-EFT is also too unspecific to provide meaningful input for fundamental constructs. 
It merely provides a theoretical space for experimental exploration.
Similar arguments apply for Stage 2, at least in general and unless going to Stage 2 was motivated by certain assumptions about candidate models.
Merely truncating the Lagrangian expansion in Eq.~\ref{eq:smeft} or focusing on the top quark or the Higgs sector does not provide a sufficiently well-defined target for BSM searches.
It is only at Stage~3, once significant deviations from the SM have been found, or are assumed to be present, that the SM-EFT can be seen as
a phenomenological model of Type-B, i.e., as about some object or system.
One may then understand SM-EFT in the same way as the early Fermi theory, i.e., as a description of some physical process that is based on vertices instead of force-mediating fields, and that could be described within some Type-A theory.
We thus find that SM-EFT, given its close relation to Type-A and B theories, in principle falls within Hartmann's hierarchical conception, but that it is not possible to interpret the first two stages as models even though they are formally expressed as theories.
The whole point of SM-EFT is to use the formal machinery of a well-confirmed theory to search for deviations, that is, to explore its limits in order to search for new physics. 
To move forward from this mixed outcome, it is helpful to recall Hartmann's distinction between the global perspective of a theory and the local understanding provided by models \citep{hartmann2001}. 
Some theories, such as SM-EFT, may fail to provide local understanding, even though typically theories unify the local understanding of the models that fall under them.

\subsection{Does SM-EFT Represent?}
\label{sub:smeftrep} 

We outlined in Sect.~\ref{sec:physrepres} the importance of representation for physicists searching for BSM effects. 
While we thus think that representation plays a significant role in the practice of particle physics, we have seen in Section ~\ref{sub:reptarget} that scholars disagree on how crucial it is in characterising modelhood. 
While representation figures among the four basic elements of the MaM approach, and in McCoy \& Massimi's application of MaM to simplified models, other practice-based approaches, among them Knuuttila's, treat representation as an asset in successful modelling practice but not as an essential precondition of being a model. 
We will discuss these two approaches in the next two sections in order to investigate whether SM-EFT represents at all stages, within the context of these approaches, and to what extent the outcome of this analysis matches the research practice in present-day elementary particle physics.\footnote{Hartmann's papers discuss the representation within the context of idealisation. Models represent some object system in an idealised fashion, and there is a variety of idealisation strategies on all four levels. Without going into further detail, we believe that his diagnosis about the representation of SM-EFT would largely agree with our conclusion concerning the MaM approach. There is nothing about which SM-EFT could be an idealisation at Stage 1.}
The key problem of our analysis will not lie in the specifics of the concept of representation applied, but in whether SM-EFT at all stages has a target that is specific enough to allow us to speak meaningfully about representation at all. 
Having a target appears to us a necessary condition for representation.

Representation in bottom-up EFTs is complex and depends on the stage.
Let us start with the Fermi theory that may be considered as a Stage-3 EFT and discuss it in the context of MaM's notion of representation. 
Even without having an idea of the larger SM framework, the original theory represented the phenomenon of beta-decay in terms of the input (down quark) and the output (electron, neutrino, and up quark) by respective wave-functions.~\footnote{In agreement with Eq.~\ref{eq:Fermi} we use modern notation.}
Similarly, one can today write down BSM approaches by using a few EFT operators, as exemplified in Sect.~\ref{sec:BphysicsConcepts}, in which each term is a representation of a field and its properties.
Such representational terms can be motivated by factual indications, by having a BSM physics idea like preons, leading to almost an identical equation, or just by a parametrization of a possible deviation from the SM without a definite physics idea like $e^+e^-\rightarrow W^+W^-$.
Thus, at Stage 3, a clear target for experimental and theoretical evaluations is identified.
This is also the case if such a target appears without any previous phenomenon or physics idea as a result of experimental investigations of deviations from the SM conducted within the context of the SM-EFT strategy that result in one or two terms of the initial expansion having non-zero coefficients.
This suffices for representation according to MaM, because models can be representative of theory or the world, or both. A model at Stage 3 either arises as representing a certain physics idea in terms of a mathematical expression in the EFT or as representing an experimentally discovered deviation in terms of a model whose description in terms of SM-EFT is still in a very preliminary form.

At Stage 2, only some broader area of interest is specified, e.g. the Higgs sector, which, 
out of pragmatic and epistemic reasons, physicists deem the most likely sector to reveal deviations from the SM.
However, there is no specific target defined and there is no representation of a BSM effect.
The space of required observations is significant and the Higgs-EFT is just a parametrization of all possible deviations one may think of, taking into account the theoretical constraints discussed in Section.~\ref{sec:eftintro}. 

This lack of a clear target and its representation is compounded in Stage~ 1. Any target of representation, if there were to be one, would be too vague to be meaningful. 
If we treated the SM-EFT or some part of it as a formal model---after all it is still a Lagrangian---it would not be clear of what it would be a model.
It is an explicit aim of the approach to remain silent on the nature of any new physics because its objective is to be sensitive for experimental deviations from the SM of whatever nature. 
Thus, it is inherent to the bottom-up approach that there is no specific target of representation in Stages 1 and 2.

To us, this indicates that SM-EFT does not exhibit all four elements identified in the MaM approach. 
The characteristics are, however, exhibited in a Stage-3 EFT, when a target of representation is specifiable, even if there is no new field posited.  
But it is clear that if there were evidence of a deviation this is precisely the next step that the physicists would take, either by devising an explicit model or a simplified model (see Section~\ref{sec:physrepres}). 
For this reason, SM-EFT as an epistemic strategy is somewhat orthogonal to the objectives of the MaM approach that starts from a given phenomenon that cannot be effectively explained by theory, if there is any, without the mediation of autonomous models. In the case of SM-EFT, the theory provides only the formal means for a rather open-ended experimental investigation. 
This might be seen in analogy to Morrison's example of the mathematical pendulum discussed in Section~\ref{sub:reptarget} that provides the theoretical means to introduce additional effects, but in contrast to this well-understood de-idealisation, particle physicists have few empirical indications of what the new physics could be like.

BSM models, which virtually all hypothesize new fields and interactions, provide a target with fairly well defined properties. 
These new fields are considered a representation of a possible real phenomenon.
Similarly, simplified models select a few entities to represent an observable phenomenon, even though it may have different physical interpretations if eventually expressed within the context of an explicit BSM model.
Simplified Models put representation centre-stage, as was emphasized by McCoy and Massimi, because one has to properly define what representation means for entities that are only partially specified. 
\citet{massimi2018} has argued more generally that such perspectival modelling avoids contradictions insofar as the same target may well be represented by incompatible models. 
She identifies four aspects of perspectival modelling: i) there is a \textit{plurality} of such models; ii) each model is only a \textit{partial} description; iii) the models are \textit{complementary}; and iv) as we noted above, they are \textit{modal} models in that they represent merely possible rather than actual states of affairs. 
A similar view cannot be applied to SM-EFT which, at Stage 1, does not represent even possible states of affairs, but merely parametrises physicists' ignorance. 
At this stage, there is no group of partial, complementary models along which one can take various incomplete perspectives on a space of possibilities. 

\subsection{The limits of a purely epistemic approach} \label{sub:epistemic}

There are also approaches that most likely would consider SM-EFT as a model at all stages. 
This is rather obvious for a syntactic account because SM-EFT is a formal model of quantum field theory. 
Considering SM-EFT as a theoretical tool for experimental searches for BSM physics renders it as an epistemic tool in the sense of Knuuttila's (\citeyear{knuuttila2011,Knuuttila2017}) artefactual account discussed in Section~\ref{sub:reptarget}.  
There can be little doubt that SM-EFT allows us to learn from data in specific ways and to obtain constraints on possible physics.
It allows one to make inferences at all stages and has a specific design that allows it to be manipulated as an artefact rather than as an abstract object that accurately represents a target.\footnote{See the characteristics mentioned in Section~\ref{sub:reptarget} above.} 
Its justification is distributed over theoretical principles and experimental strategies---while traditional particle physics models are justified by their resolving a certain problem of the SM or by introducing some idea of new physics.
The widespread use of SM-EFT in the community documented in Section~\ref{sec:data} and the existence of a longer tradition of effective theories \citep{wells2012} indicate that the method is indeed `culturally established'. 
The bottom-up EFTs are based on the same experimental and theoretical techniques as studies of the SM and simplified models and explicit models. 

At each stage, the bottom-up SM-EFT approach is described by a Lagrangian 
obeying the principles also used for the SM and the constraints on the EFT parameters are obtained in the same experiments that test BSM models using the same experimental techniques. 
However, there are differences of strategy: theoretical and experimental studies of bottom-up EFT approaches emphasise deviations, whereas studies of BSM models mostly look for the footprint of a new particle against smooth background.
Thus, for epistemic accounts, there seems to be no principal difference between explicit models and the bottom-up approach that eventually aims to arrive at them. 
Since in epistemic accounts of models, representation is a welcome add-on and its absence does not automatically disqualify something from being a model, the bottom-up EFTs, including Stages 1 and 2, would be classified as models.

Fair enough, but the problem we see in Knuuttila's epistemic account is that it does not reflect a key element of the scientific practice of most particle physicists. 
While Knuuttila considers representations and its tools as all of a piece, the SM-EFT strategy puts an extra value on the goal of representation. 
As outlined in Sect.~\ref{sec:physrepres}, physicists expect a model to have a well defined target of representation. 
While, as discussed in Sect.~\ref{sub:smeftrep} such a target exists at Stage 3, it is absent at Stages 1 and 2, as exemplified in Sect.~\ref{sec:BphysicsConcepts}. 
For particle physicists, the Stages 1 and 2 are just (if at all) stepping stones towards a representational model of BSM physics.
If induced by experimental findings, the transition between Stages 2 and 3 would signify a major breakthrough.

\subsection{Lessons for the Philosophy of EFT} \label{sub:EFT}

Our analysis of how three philosophical accounts of models address bottom-up EFT strategies, in particular SM-EFT, puts Hartmann's main question, the relationship between model and theory in assessing EFT, back into the focus and gives it a somewhat surprising twist.
SM-EFT at Stage 1 can fail to be a model, at least on the MaM account, and yet still qualify as a theory, while it may have acquired, at Stage 3, the characteristics of a model.
SM-EFT is a theoretical tool that helps experimental physicists to parametrize their constraints or observations of deviations from the SM. 
Such searches may or may not be guided by any explicit models or aspects of models.

A bottom-up EFT is certainly not guided by a model or a theory in the usual sense. But our case study clearly shows that there exists a role for theory in scientific practice without the mediation of models. 
This is somewhat at odds both with the general idea of models as mediators between theory and data, and the hierarchical approach of Hartmann and advocates of the semantic approach. 
To be sure, theory in this case does not predict, unify, or explain. 
It is primarily a tool for the parametrization of experimental results that involve research motives that are not connected to specific models or theories.

It seems to us that this differentiation also bears important lessons for the present debates about EFT. 
As mentioned in Sect.~\ref{PhilEFT}, most philosophers understand bottom-up EFTs either as ways to discover a new fundamental level in the tower of theories  or as a way to study the limits of an already specified EFT and support its realist interpretation through a robustness argument. 
If there are any effects of unknown physics at higher energy, they can be absorbed into the parameters of the EFT on the basis of experiments conducted on the scale of its validity. 
In doing so, some interpreters consider the limit of validity, the cut-off, as an element of physical reality. When physicists are applying SM-EFT, realism works the opposite way. 
The SM---or in Stage 2: certain sectors of it---are assumed to be real and its parameters are locked in, which allows physicists to look for deviations from those values of whatever kind and without any commitments about their nature. 
They are looking for constraints on BSM physics or some target of further analysis---for some thing to model. 
It seems to us that philosophically this goal to obtain a target and eventually a model that represents new physics is more modest in kind than the realist discourse that present discussions about EFT are embedded in and physicists applying SM-EFT may remain agnostic about realism vs. instrumentalism or reduction vs. emergence. 
This extends to their attitude about the cut-off. It has to allow potential new physics to reveal itself, but not enough to dominate the physics at the low energy level.
Emphasizing this more modest goal, we do not deny that there are relations between the issues of representation and realism as regards EFT. \footnote{We thank an anonymous referee for pointing this out to us.}
To give an example, Williams' \citeyear{williams2018} interpretation of effective realism in terms robustness (of the fields and parameters of the standard model) can well be understood as the goal of the entire SM-EFT strategy in the sense that while the parameters of the SM are considered robust and fixed from the start, physicists try to find a robust description of BSM physics in terms of one or a few coefficients in the Stage 3 EFT. After all, robustness is also a concept describing experimental data.

\section{Conclusion}

In this paper, we have examined one of the most interesting philosophical aspects of a recent change in particle physics, the turn towards model-independent and experiment-focused research at the LHC, focusing on the example of the SM-EFT. 
We have found that understanding its relation to models and theories not only depends on the philosophical understanding of models but also on the stages of this research practice. 
The generic SM-EFT at Stage 1 lacks a target of representation and as we have argued, to be a model is to be a model of something and so it lacks a key characteristic of being a model. 
At Stage 2, one focuses the SM-EFT on specific experimental sectors in a pragmatic way makes the SM-EFT experimentally and phenomenologically useful, but this does not add a sufficiently well-defined target. 
Additional theoretical input or experimental evidence of a deviation from the SM can precisify a target of representation and the SM-EFT would then autonomy from the SM. 
It is only in this last stage (Stage 3) that it takes on the characteristics of a model.
We take this as a reason---motivated not primarily by philosophical theory but by existing research practices---not to take a view on models that is overly permissive: it blurs the lines between the different stages of the use of SM-EFT and glosses over particle physicists' methodological aims of undertaking the bottom-up approach in the first place. 
Or put otherwise, not every step in a research strategy that successfully combines theoretical tools with experimental investigations and eventually leads to a new scientific model should be considered as a model itself.

Another philosophical result of our case study is that whether something is a model or not, may change within the different stages of an epistemic strategy and also depend on the status of empirical knowledge available or assumed.
Even though we have described the SM-EFT strategy as going through different stages, they need not occur in real time or in sequence. 
For example, the SM-EFT framework can be used at the same time in one sector of measurements as a tool to describe constraints on deviations from the SM, and in a different sector of measurements as a Stage~3 EFT model of observed deviations from the SM.
It is also possible that a model that results as Stage 3 of a successful SM-EFT strategy in a bottom-up approach could also have been proposed independently in a top-down approach on the basis of some physics idea.
But, as shown, it is also possible that empirical evidence, in terms of deviations from the SM, prompts physicists to move through the three stages. 
Our point is just that modelhood can change within the context of an epistemic strategy without the prior application of specific modelling assumptions, but through empirical evidence gained in its wake.

\section{Acknowledgements}
We thank the members of our research group and two anonymous referees for their targeted criticism and many helpful suggestions. This research was funded by the German Research Foundation (DFG grant FOR 2063). 

\bibliography{smeference}
\bibliographystyle{plainnat}

\end{document}